\documentclass[twocolumn, prc, amssymb, superscriptaddress, aps,
preprintnumbers,amsmath,floatfix]{revtex4}

\usepackage{multirow}
\usepackage{isotope}
\usepackage{supertabular}
\usepackage{graphicx,amsfonts,color,epsfig}
\usepackage{amsmath}
\usepackage{amssymb}
\usepackage{url}
\usepackage{xcolor}
\usepackage{tikz}
\usepackage{xcolor}
\definecolor{Bxcol}{RGB}{90,10,0}

\newcommand{\ket}[1]{|#1\rangle}
\newcommand{\expval}[1]{\langle#1\rangle}
\newcommand{\vect}{\mathbf}

\newcommand{\elmx}[3]{\langle #1|#2|#3\rangle}

\begin{document}


\title{Skyrme-Hartree-Fock-BCS approach to $^{229m}$Th and
  neighboring nuclei}


\author{Nikolay \surname{Minkov}}
\email{nminkov@inrne.bas.bg} \affiliation{Institute for Nuclear Research and
Nuclear Energy, Bulgarian Academy of Sciences, Tzarigrad Road 72, BG-1784
Sofia, Bulgaria}

\author{Adriana P\'alffy}
\email{adriana.palffy-buss@uni-wuerzburg.de}
\affiliation{University of W\"urzburg, Institute of Theoretical Physics and Astrophysics, Am Hubland, 97074 W\"urzburg, Germany}

\author{Philippe Quentin}
\email{quentin@cenbg.in2p3.fr}
\affiliation{LP2i Bordeaux, UMR 5797, Universit\'e de Bordeaux, CNRS, F-33170,
  Gradignan, France}
\author{Ludovic Bonneau}
\email{bonneau@lp2ib.in2p3.fr}
\affiliation{LP2i Bordeaux, UMR 5797, Universit\'e de Bordeaux, CNRS, F-33170,
  Gradignan, France}


\date{\today}

\begin{abstract}
The  microscopic origin of the $^{229m}$Th isomer and its possible manifestation also in neighboring nuclei is explored within a selfconsistent Skyrme Hartree-Fock plus Bardeen-Cooper-Schrieffer  approach. By using the well established SIII Skyrme parametrization, without any special adjustments related to low-energy  isomer, the single-particle spectrum provided by the model reproduces the correct isomer $K^{\pi}=3/2^{+}$ spin and parity, and the relative proximity of the isomeric state to the $K^{\pi}=5/2^{+}$ ground state, yet on the keV scale. We show that this approach may provide microscopic estimates for some related observables, such as the quadrupole and octupole moments, deformation parameters as well as  magnetic dipole moments. Its ability to provide a prediction for the M1 isomer transition probability is discussed. The obtained $^{229}$Th single-particle structure is compared with that provided by calculations in neighbouring actinide isotopes and isotones, allowing us to assess the more general role of the considered mechanism for the formation of low-energy isomers.
\end{abstract}


\maketitle

\section{Introduction}
\label{intro}

A unique nuclear isomer with energy of only 8~eV has become recently the subject of intense investigation in atomic, nuclear and laser physics and metrology. The $^{229m}$Th isomer  is so far an exceptional  case of nuclear excitation optically accessible by suitable laser sources in the vacuum-ultraviolet (VUV). This promises a new frequency standard \cite{Peik_Clock_2003,Campbell_Clock_2012,Peik_Clock_2015}, and means to probe the variation of fundamental constants \cite{flambaum_2006,Fadeev_2020,peik2021nuclear} or search for light dark matter \cite{peik2021nuclear,tsai2023direct} in unprecedented ways. $^{229}$Th has a ground state (GS) spin and parity 5/2$^+$, an isomeric state (IS) spin and parity 3/2$^+$ and the transition between the two states represents a magnetic dipole $M1$ and electric quadrupole $E2$ mixture. The experimental determination of the transition energy has come a long way in the past decades \cite{Beck_78eV_2007,Beck_78eV_2007_corrected,Seiferle_EnTh229m_2019, Yamaguchi_EnTh229m_2019,Sikorsky2020,kraemer2023observation,PRL2024}, involving observation of both  the internal conversion \cite{Wense_Nature_2016,Seiferle_PRL_2017} and the radiative decay channels  \cite{kraemer2023observation,PRL2024,elwell2024laser}.
The breakthrough for the design of the nuclear clock occurred this very year (2024), with three reported experiments which succeeded in direct laser driving of the isomeric transition in Th-doped VUV transparent crystals  \cite{PRL2024,elwell2024laser,JunYe2024}. The experimental accuracy for the nuclear transition frequency increased from GHz \cite{PRL2024,elwell2024laser} to kHz in the most recent experiment using a VUV frequency comb \cite{JunYe2024}.

From nuclear structure perspective, early theoretical estimates for the isomer reduced transition probability $B(M1)$  and the isomer magnetic moment  $\mu_{\mbox{\scriptsize IS}}$ were made in Refs.~\cite{Dyk98,Tkalya15} using the Alaga branching ratios \cite{Alaga55} and the Nilsson model \cite{Nilsson1955}. Alternative predictions for the $B(M1)$ and $B(E2)$ reduced transition probabilities for the $3/2^{+}\rightarrow 5/2^{+}$ isomer transition have been made in Refs.~\cite{Gulda02,Ruch06} using the quasiparticle-plus-phonon model \cite{Sol76} in an overall description of the $^{229}$Th spectrum without particular consideration of the isomer properties. Furthermore, microscopic calculations involving the $3/2^{+}$ isomer energy have been performed within the Relativistic Mean Field (RMF) theory \cite{JPG07_He,NPA08_He} and the Hartree-Fock-Bogoliubov (HFB) approach \cite{PRC09_Litvinova_HFB} as a part of the efforts to determine the isomer decay rate sensitivity to possible temporal variations of the fine structure constant $\alpha$, without particular focus on the origin and formation mechanism of the isomer.

The $^{229m}$Th isomer formation mechanism was recently described within a model approach providing an explanation for the appearance of the extremely low-energy state as the effect of a very fine interplay between collective quadrupole-octupole degrees of freedom and the single-particle (s.p.) motion of the odd neutron in $^{229}$Th \cite{Minkov_Palffy_PRL_2017,Minkov_Palffy_PRL_2019,Minkov_Palffy_PRC_2021}. The IS was considered as the bandhead of a quasi-parity-doublet spectrum built on a $K^{\pi}=3/2^{+}$ quasiparticle (q.p.) excitation (3/2[631] s.p. orbital) which is coupled via Coriolis mixing to the $K^{\pi}=5/2^{+}$ (5/2[633] orbital) GS. Here, $K$ and $\pi$ refer to the projection of the total nuclear angular momentum $\hat{I}$ on the body-fixed principal symmetry axis and the parity, respectively, while the notation $K[Nn_z\Lambda]$ involves the usual asymptotic Nilsson quantum numbers \cite{NR1995}, explained in more detail below. The approach combines the features of a collective (coherent) quadrupole-octupole model (CQOM) \cite{b2b3mod,b2b3odd,NM13,MDSSL12,MDDSLS13} and deformed shell model (DSM) \cite{qocsmod} with pairing correlations of Bardeen-Cooper-Schrieffer (BCS) type. The overall CQOM-DSM model approach provides theoretical predictions for the $B(M1)$ and $B(E2)$ isomer decay rates \cite{Minkov_Palffy_PRL_2017} and GS and IS magnetic moments \cite{Minkov_Palffy_PRL_2019} in $^{229}$Th. Also, it allows one to make a detailed assessment of the shape-deformation and intrinsic-motion conditions, including the quasi-degeneracy of the two s.p. orbitals, which determine the isomer-formation mechanism and the resulting energy and electromagnetic properties of the metastable state \cite{Minkov_Palffy_PRC_2021}.

The relevance of the CQOM-DSM approach in the explanation of key $^{229m}$Th properties raises the question about the validity of the proposed isomer-formation mechanism on the deeper microscopic mean-field level. It is therefore important to understand how the deformation and intrinsic effective interactions determine the isomer-formation conditions in the solution of the many-body problem for $^{229}$Th. More specifically, it is interesting to examine how deformations and effective interactions govern the intrinsic states which could be responsible for the isomer excitation, the related energy-levels proximity or quasi-degeneracy, the overall energy and electromagnetic characteristics of the nucleus. The same question naturally extends to other nuclei in the same mass region requiring to check how the isomer-formation conditions evolve with the changing intrinsic structure.

The purpose of this work is to clarify the above questions within the
framework of the Skyrme energy-density functional (EDF) approach,
including BCS pairing correlations with selfconsistent blocking. To
solve this task we employ a Hartree-Fock-BCS (HFBCS) approximation
with time-odd terms and seniority pairing correlations as formulated
in Ref.~\cite{PRC15_hfbcs}. As a basic Skyrme EDF option we take the
SIII Skyrme parametrization \cite{Beiner75_SIII}. The time-odd terms,
taken in a ``minimal'' scheme with spin and current vector fields,
as defined in Ref.~\cite{PRC15_hfbcs} and recalled here in
Appendix~A for self-consistency, allow one to take into account the polarization effect of the unpaired nucleon on the even-even core resulting in time-reversal symmetry breaking, and hence a suppression of Kramers degeneracy. The algorithm is elaborated for calculations without imposed reflection symmetry, still keeping the axial symmetry, which allows for possible solutions with non-zero ``pear-shape'' octupole deformation. The selfconsistent equations are solved by diagonalization of the Hartree–Fock (HF) Hamiltonian in an axially-symmetric deformed harmonic oscillator basis \cite{Vautherin73}. The approach is especially suitable for nuclei in the actinide mass region known to exhibit axial quadrupole or axial quadrupole-octupole deformations.

Some basic explanations of the approach and its application for description of magnetic moments and spin-gyromagnetic quenching in odd-mass nuclei have been presented in Ref.~\cite{PRC15_hfbcs}. In Refs.~\cite{BJP19_hfbcs_isomers,BJP21_hfbcs_isomers,PRC22_hfbcs_isomers,PRC24_hfbcs_isomers} the model was applied to $K$-isomeric excitations in heavy even-even nuclei. Its capability to handle octupole deformation was demonstrated in Ref.~\cite{BJP19_hfbcs_isomers}. All these HFBCS applications with the SIII force  show that the approach allows a reasonably good description of the spectroscopic properties of interest. We remark that, owing to the BCS treatment of pairing and the Slater approximation used for the exchange Coulomb terms, the present study is expected to share similarities with the HFB calculations of Ref.~\cite{PRC09_Litvinova_HFB}, but brings new insights because of the breaking of time-reversal symmetry in the one-quasiparticle intrinsic states inherent to selfconsistent blocking. The other essential feature of the present study is the consideration of the reflection-asymmetric (octupole) deformation, missing in the result of Ref.~\cite{PRC09_Litvinova_HFB}, which turns out to play a crucial role in the understanding of the $^{229m}$Th isomer properties.

It should be noted here that presently there is a large variety of nuclear EDF approaches with developed high-performance computer codes which can be applied to the $^{229m}$Th problem. For instance, we refer here to the HFODD solver for the HF and HFB problem in Cartesian deformed harmonic-oscillator basis \cite{JPG21_Dobaczewski}, the Skyrme HFBCS Ev8 \cite{cpc15_Ryssens} and HFB Sky3D \cite{cpc22_Sky3D} codes realized in three-dimensional coordinate space and many others (see also further references in these articles). All these state-of-the-art approaches can provide an advanced symmetry-unrestricted treatment of $^{229m}$Th with high-computation efficiency and numerical accuracy, use of a large variety of effective interactions, and many other features.  At the same time, we consider that the present Skyrme HFBCS approach, with the extensively tested SIII interaction, is capable to provide a relatively simple and easily understandable solution. It adequately incorporates the key features of an odd-mass nucleus with core polarization, pairing correlations, axial quadrupole-octupole deformation and the resulting energy and electromagnetic properties.

We apply the Skyrme HFBCS approach with the characteristics described above to the particular study of the $^{229}$Th ground-state and the lowest q.p. excitations, trying to infer the possible formation mechanism of the $3/2^{+}$ isomer and its spectroscopic properties. It is clear that the accuracy level of the nuclear self-consistent approach is far away from the eV precision needed to reproduce the measured isomer energy. Nevertheless, the thorough examination of the $^{229m}$Th HFBCS solution under different model conditions is of an utmost importance to understand the microscopic origin of the isomer-formation mechanism and to follow its possible evolution in other nuclei as well. Such a study requires a fine evaluation of some basic conditions, including the core polarization induced by the unpaired neutron, the pairing correlations and the effect of reflection-asymmetric (octupole) shape, in much deeper details than usually considered in the global application of the standard many-body approaches. To this end we have structured our study in separate steps allowing us to assess the effect of adding the single nucleon to the even-even core and imposing or not reflection symmetry, on the formation of low energy q.p. excitations including the extremely low-lying isomer state.

Our step-wise procedure goes as follows. We start with the solution for the even-even core nucleus $^{228}$Th for which the s.p. orbitals around the Fermi level are analysed both in the reflection-symmetric and octupole cases, with a focus on the relative position of the $3/2^{+}$ and $5/2^{+}$ orbitals. This allows us to gain a deeper understanding of the structure changes occurring after the unpaired neutron is added in $^{229}$Th. (In principle, one can equally well consider $^{230}$Th as a core nucleus from which a neutron is removed.) We then proceed with the HFBCS iterative process with selfconsistent blocking using the same pairing constants and basis parameters of the GS solutions of the core, and find the corresponding $^{229}$Th solutions, with and without reflection symmetry, for the $5/2^{+}$, $3/2^{+}$ and a few other neighboring q.p. excitations. This allows us to assess the ability of the HFBCS algorithm to handle the proximity of the $3/2^{+}$ excitation to the $5/2^{+}$ GS under the different shape conditions. For each state we obtain model predictions for a number of structure quantities---deformations, electric and magnetic moments and gyromagnetic factors, with some of them being compared with values obtained in the CQOM-DSM approach. We note that the magnetic dipole moments are of special importance since their values incorporate the underlying electromagnetic characteristics of the considered states and in particular the $^{229m}$Th isomer decay properties.

The calculations are extended to the neighboring $^{227}$Th and $^{231}$Th isotopes and $^{227}$Ra and $^{231}$U isotones, taking $^{226}$Th, $^{230}$Th and $^{226}$Ra, $^{230}$U, respectively, as initial even-even cores for the calculation. This allows us to examine the ``migration'' of the $^{229m}$Th isomer-forming s.p. orbitals and their close neighbors in the vicinity of $^{229}$Th. The results outline quite a detailed picture for the evolution of the intrinsic conditions which may govern the emergence of low-energy metastable states in the considered actinide mass region.

The paper is structured as follows. In Sec.~\ref{adapt} we apply the HFBCS approach to the core nucleus $^{228}$Th. Subsequently, in Sec.~\ref{th229calc} the elaborated prescription is applied to $^{229}$Th for the two cases of reflection-symmetry and octupole deformation. The results are analysed and we discuss  the capability of the used effective interactions (both in the particle-hole and particle-particle/hole-hole channels)  to provide conditions appropriate for isomer formation. We extend our  study to the neighboring nuclei  in Sec.~\ref{neighb} outlining the evolution of the overall low-energy excitation conditions around~$^{229}$Th. The paper concludes in Sec.~\ref{over} with a discussion and remarks on further perspectives of our approach for the study of~$^{229m}$Th and similar phenomena in other nuclei.

\section{HFBCS approach to the Thorium problem. The $^{228}$Th core.}
\label{adapt}

\subsection{HFBCS solution with and without reflection symmetry}

The HFBCS algorithm applied with reflection symmetry provides a solution with axial quadrupole deformation. The basis is symmetrized with respect to the $z$-axis origin and the variation procedure starts with an initial Woods-Saxon  potential with deformation defined by the quadrupole parameter $\beta_{2}=0.2$. The converged result includes the obtained quadrupole moment and, for the odd mass nucleus, the magnetic dipole moment and gyromagnetic factor. In this case the intrinsic parity is conserved and the s.p. wave functions possess a definite parity.

When the reflection symmetry is no longer imposed, one is often confronted with convergence difficulties because of shifts in the center-of-mass  position on the symmetry axis during the iteration procedure. This requires a specific center-of-mass constraint to fix its position at the origin of the intrinsic frame. In addition, to allow the HFBCS algorithm to find a potential octupole solution, we start the calculation with an initial mean field generated by a Wood-Saxon potential with the octupole deformation parameter value of $\beta_3=0.1$ in addition to the quadrupole $\beta_2=0.2$ deformation. We consider that the model indicates the presence of octupole shape if the solution is obtained with non-zero octupole moment and $\beta_3$ deformation and the resulting total energy is lower compared to that in the reflection-symmetric (pure quadrupole) solution. This solution corresponds to axial quadrupole-octupole deformation, while the s.p. wave functions appear with mixed parity and can be only characterized by the average (expectation) value $\langle\pi\rangle$ of the parity operator calculated in the given state.

\subsection{Axially-deformed harmonic oscillator basis}

The single-particle HF Hamiltonian is diagonalized in the axially-deformed harmonic oscillator (ADHO) basis~\cite{Vautherin73}. This basis is characterized by the oscillator frequencies along the symmetry axis (chosen to be the $z$ axis), $\omega_z$, and in a direction perpendicular to the $z$ axis, $\omega_{\perp}$. These two frequencies define $\omega_0$ as an equivalent (volume conserving) frequency by $\omega_0^3 = \omega_z {\omega_{\perp}}^2$. The ADHO basis is conveniently referred to through the spherical equivalent oscillator constant $b=\sqrt{m\omega_0 /\hbar}$, where $m$ is the mean nucleon mass, and the deformation parameter $q=\omega_{\perp}/\omega_{z}$.

In practice the ADHO basis is truncated as explained in Ref.~\cite{Floc73}. The basis states are eigenvectors of the axially-deformed, two-dimensional harmonic-oscillator Hamiltonian and carry the Nilsson quantum numbers $\ket{Nn_z\Lambda\Sigma}$ where $n_z$ denotes the phonon number along the symmetry axis, $N=n_z+n_{\perp}$ is the total phonon number, with $n_{\perp}$ the phonon number in a direction perpendicular to the symmetry axis, $\Lambda$ and $\Sigma$ are the projections on the symmetry axis of the orbital and spin angular momenta, respectively. One has $K=\Lambda +\Sigma$. The states $\ket{Nn_z\Lambda\Sigma}$ retained in the truncated basis satisfy the condition $\omega_z (n_z + 1/2) + \omega_{\perp} (n_{\perp} + 1) \leqslant \omega_0 (N_0 + 3/2)$, where the basis truncation parameter $N_0$ is such that $N_0+1$ represents the number of retained spherical shells when $q=0$ and takes here the value $N_0 = 16$.

Because of the finiteness of the s.p. basis, we have to optimize the $b$ and $q$ parameters by minimizing the total energy for the ground-state solution (including the pairing energy). For the $^{228}$Th core nucleus we find $b=0.480$ and $q=1.20$.

Finally, all numerical integrations, in particular to compute the matrix elements of the Hartree-Fock Hamiltonian, are performed by Gauss-Hermite (over the $z$ coordinate) and Gauss-Laguerre (over the radial coordinate) quadratures with 30 and 15 points, respectively.

\subsection{Determination of the BCS pairing strengths}

In our approximate treatment, we define different constant neutron and proton averaged matrix elements determining the strength of the BCS pairing correlations up to 6 MeV above the chemical potentials with a smoothing factor defined by a width of 0.2 MeV (see Ref.~\cite{Pillet2002} for details). For a charge state $\tau=n,p$ ($n$ - neutrons, $p$ - protons), their dependence on the corresponding fermion numbers $N_{\tau}$ is parametrized as $V_{\tau} = -G_{\tau}/(11 + N_{\tau})$ (see Ref.~\cite{Bonche1985}). In a recent paper \cite{PRC19GnGpfit} some of the present authors have shown that
in well deformed nuclei, the determination of
the two  strength parameters $(G_n , G_p)$ above can be almost equivalently performed either by fitting three-point odd-even mass differences $\Delta^{(3)}$ (around odd nuclei) or the energy of the first $2^{+}_{1}$ level in even-even nuclei. In  the latter case the
 $2^{+}_{1}$ energy is related to the calculated moment of inertia  $\mathfrak{J}$ within the simple quantal rotor approximation $E(2_{1}^{+})=\frac{\hbar^{2}}{2\mathfrak{J}}[2(2+1)]=3\hbar^{2}/\mathfrak{J}$.

In view of the easier numerical task associated with computing
$\mathfrak{J}$ with respect to $\Delta^{(3)}$, we chose to perform our
fit within the second approach. We note that the moments of inertia of even-even nuclei are calculated within the usual Inglis-Belyaev approach~\cite{Inglis54,Inglis55,Belyaev61} supplemented by so-called self-consistent Thouless-Valatin corrections~\cite{Yuldashbaeva1999}. The enhancement coefficient (in practice a $1.32$ multiplication factor) to take care of these corrections has been taken from Ref.~\cite{Libert99}
where it has been determined upon using the D1S Gogny  interaction and found in many instances (see, e.g., \cite{PRC22_hfbcs_isomers}) to be valid also for selfconsistent solutions using the SIII interaction. It is thus clear that our calculated moments of inertia depend on three important approximations (pure rotor dynamics, adiabaticity and approximate Thouless-Valatin corrections).

Keeping the above in mind, upon fitting the pairing strengths on the experimental $E(2_{1}^{+})$ value  of 57.8 keV \cite{ensdf} for the $^{228}$Th ground state solution we obtain $G_n=16.0$~MeV and $G_p=14.4$~MeV  corresponding to a theoretical energy  value of $E^*=58.4$ keV. These values for the pairing strength parameters are close to the results of a systematic fit in  Ref.~\cite{PRC22_hfbcs_isomers} which yielded $G_n = 15.80$ MeV and $G_p = 14.23$~MeV.

\begin{table*}[ht]
\begin{center}
\caption{Total energy ($E_{\mbox{\scriptsize tot}}$), first $2^{+}_{1}$
excitation energy approximated by $E_{\rm th}^*(2_1^+) =
3\hbar^2/\mathfrak{J}_{th}$, neutron ($\Delta_{n}$) and proton
($\Delta_{p}$) pairing gaps, mass and charge quadrupole moments
($Q_{20}^m$ and $Q_{20}^p$), and quadrupole deformations
($\beta_{2}^{m}$ and $\beta_{2}^{p}$), mass octupole moment
($Q_{30}^{m}$) and octupole deformation ($\beta_{3}^{m}$) obtained
for the ground state of $^{228}$Th in the reflection-symmetric and
reflection-unconstrained HFBCS solutions with different sets of
pairing strengths $G_n$ and $G_p$ (see text for further explanations).}
\label{tab:th228hfbcs}
\bigskip
\tabcolsep=3pt
\begin{tabular}{ccccccccccccc}
\hline\hline
\noalign{\smallskip}
$G_n$ & $G_p$ &$E_{\mbox{\scriptsize tot}}$ & $E_{\rm th}^*(2_1^+)$
&$\Delta_{n}$&$\Delta_{p}$& $Q_{20}^m$ & $\beta_{2}^{m}$ & $Q_{20}^p$ & $\beta_{2}^{p}$ &$Q_{30}^{m}$ & $\beta_{3}^{m}$\\
(MeV)  & (MeV)  &  (MeV) &  (keV) & (MeV) & (MeV) &    (fm$^2$)  &  &    (fm$^2$)  & & ($b^{3/2}$)&       \\
\hline
\noalign{\smallskip}
&\multicolumn{10}{c}{ Reflection-symmetric solution}&\\
16.0 & 14.4 & $-$1737.900 & 58.4 & 0.867 & 0.888 &2057.042 & 0.210 &818.514 & 0.212 & -- & --
\medskip\\
&\multicolumn{10}{c}{ Reflection-asymmetric solution}&\\
16.0 &14.4 & $-$1738.4118 & 49.3 & 0.826 & 0.765 & 1967.5 & 0.200 & 783.085& 0.202& 2.792 & 0.130 \\

16.5 &14.9 & $-$1738.7811 & 58.0 & 0.952 & 0.879 & 1954.1 & 0.199 & 777.504& 0.201& 2.659 & 0.124 \\

\hline\hline
\end{tabular}
\end{center}
\end{table*}

\subsection{Exploring the $^{228}$Th core}
\label{th228core}

We have performed HFBCS calculations for the $^{228}$Th core using the parameters discussed above with and without imposed  reflection symmetry.
The results are given in Table~\ref{tab:th228hfbcs}. They include the
theoretical total energy and $2^{+}_{1}$ excitation energy approximated by $E_{\rm th}^*(2_1^+) = 3\hbar^2/\mathfrak{J}_{th}$,
as well as other characteristics of the solutions, namely neutron and
proton pairing gaps $\Delta_{n}$ and $\Delta_{p}$, mass and charge
quadrupole moments $Q_{20}^m$ and $Q_{20}^p$,  respectively, and the
corresponding quadrupole deformation parameters $\beta_{2}^{m}$ and
$\beta_{2}^{p}$. In addition, the mass octupole moment $Q_{30}^{m}$ and
octupole deformation $\beta_{3}^{m}$ are given for the
reflection-unconstrained (asymmetric) solution. The relevant definitions and
relations connecting the quadrupole and octupole moments calculated in
the code and the correspondingly deduced deformation parameters are
given in Appendix~B.

The following comments can be made upon investigating the results given in Table~\ref{tab:th228hfbcs}:

(i) The pairing strengths fitted in the reflection symmetric solution lead to an underestimation of the experimental $2^{+}_{1}$ excitation energy in the reflection-unconstrained calculation. The last line of Table~\ref{tab:th228hfbcs} shows the results obtained after a re-adjustment of the pairing strengths to more closely match the theoretical value to the experimental one $E(2_{1}^{+}) = 57.8$~keV, which yields $G_n=16.5$~MeV and $G_p=14.9$~MeV (keeping a constant ratio $G_p/G_n = 0.9$).

(ii) In the reflection-asymmetric solution the total energy is by 0.5~MeV lower than that in the reflection-symmetric solution for the initial  pairing strengths $G_n=16.0$~MeV and $G_p=14.4$~MeV. The former solution is characterized by a non-zero octupole moment $Q_{30}^{m}$ and the corresponding octupole deformation parameter is about $\beta_{3}^{m} =0.12-0.13$. This result is in agreement with earlier calculations of $^{228}$Th ground-state properties within the self-consistent relativistic Hartree-Bogoliubov model of Ref.~\cite{Nomura2014}.

\begin{figure}[ht]
\centering
\includegraphics[width=8cm]{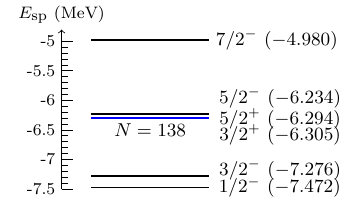}
\caption{Neutron single-particle levels near the Fermi level of
  $^{228}$Th obtained in reflection-symmetric Skyrme-SIII HFBCS
  calculation with $G_n=16.0$ MeV and $G_p=14.4$ MeV. Each level
  denoted by $K^{\pi}$ is Kramers degenerate and its energy (in MeV)
  is given in brackets. The lines depicting the $3/2^{+}$ and $5/2^{+}$
  orbitals, drawn in red and blue, respectively, are degenerate and thus indistinguishable
  in the given scale. Among them, the lower lying $3/2^{+}$ orbital is the Fermi level. }
\label{fig_228Th_sprefsym}
\end{figure}

\begin{figure*}[ht]
\centering
\includegraphics[width=\textwidth]{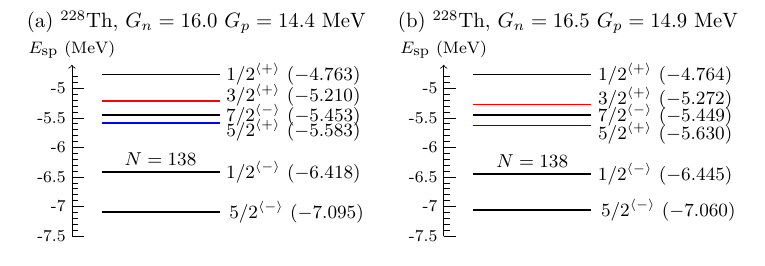}
\caption{Neutron single-particle levels near the Fermi level of
$^{228}$Th obtained in reflection-unconstrained Skyrme-SIII HFBCS
calculation with (a) $G_n=16.0$, $G_p=14.4$ MeV and (b) $G_n=16.5$,
$G_p=14.9$ MeV. Each level denoted by $K^{\langle\pi\rangle}$ is Kramers
degenerate and its energy (in MeV) is given in parenthesis. The notations  $\langle \pm \rangle$ correspond to the sign of the average parity $\langle\pi\rangle$. The lines depicting the $3/2^{+}$ and $5/2^{+}$ orbitals are drawn in red and
blue, respectively.}
\label{fig_228Th_spoct}
\end{figure*}

Having obtained the solutions in Table \ref{tab:th228hfbcs}, we can analyze the s.p. spectrum of the core nucleus $^{228}$Th in its ground state. More precisely, we inspect the positions of the neutron orbitals near the Fermi level. Because of the good-rotor character of the $^{228}$Th nucleus and its neighbors, we may expect that the positions of the $5/2^{+}$ and  $3/2^{+}$ orbitals carry information about the formation of the ground and isomeric states in the neighboring odd-mass nucleus $^{229}$Th after a neutron is added to the core. In Fig.~\ref{fig_228Th_sprefsym} the s.p. levels of $^{228}$Th obtained in the reflection-symmetric case with the pairing strengths $G_n=16.0$ MeV and $G_p=14.4$ MeV are displayed. On the MeV energy scale of the figure, the two orbitals $5/2^{+}$ and $3/2^{+}$ are found to be virtually degenerate. The $3/2^{+}$ orbital stays on the Fermi level at neutron number $N=138$, while the $5/2^{+}$ orbital is located only 11 keV above it. In addition, in Fig.~\ref{fig_228Th_sprefsym} we see a $5/2^{-}$ orbital appearing 60 keV above the $3/2^+$ and $5/2^+$ orbitals. Its presence could be related with the appearance of octupole deformation in the ground-state shape of $^{228}$Th and may play a role in forming the structure of the $^{229}$Th spectrum related to octupole deformation. We also notice the other neighboring orbitals, $7/2^{-}$ above and $3/2^{-}$ and $1/2^{-}$ below the Fermi level, which, in addition, may play a role in the formation of s.p. excitations in the nuclei around $^{228}$Th. It is interesting to remark that the same orbitals are identified in the DSM calculations for $^{229}$Th performed in the space of quadrupole $\beta_{2}$ and octupole $\beta_{3}$ deformations (see Fig.~1 in Ref.~\cite{Minkov_Palffy_PRC_2021}). It is suggested there that all these orbitals may cross each other in certain deformation ranges and, therefore, it can be expected that their interaction will influence the related structure properties of the nuclei in the region of $^{229}$Th.

The s.p. spectra associated with orbitals from the reflection-asymmetric solutions of Table~\ref{tab:th228hfbcs}, i.e., the case with octupole deformation, are shown in Fig.~\ref{fig_228Th_spoct}. The notation $\langle + \rangle$ ($\langle - \rangle$) in the s.p. labels denotes a positive (negative) expectation value of the parity operator. We notice that the $3/2^{\langle + \rangle}$ orbital now appears about 360 keV above the $5/2^{\langle + \rangle}$ level.  This energy difference is significantly larger than the corresponding energy difference of 11 keV in the reflection-symmetric case presented in Fig.~\ref{fig_228Th_sprefsym}. Also, the $5/2^{\langle - \rangle}$ orbital now appears considerably lower in energy than the $5/2^-$ level.

To understand this behaviour, it is worth recalling that the reflection asymmetric shape of the ground state of $^{228}$Th shows us that the intrinsic parity breaking in the HFBCS solution for this nucleus is energetically favorable. This is manifested by parity mixing in the s.p. states, which is according to Refs.~\cite{Gustafsson71,Johansson61} especially large in pairs of states whose dominant Nilsson configurations are of the form $K^{\pi}[Nn_z\Lambda]$ and $K^{-\pi}[N'n'_z\Lambda]$ where $|N'-N|=1,$ 3 and $|n'_z-n_z|=1$, 3, with $\pi =(-1)^{N}$. These selection rules are associated with the octupole-deformed central mean field, which is dominant as compared to the spin-orbit field. For example in the solution obtained with $G_n=16.0$~MeV and $G_p=14.4$ MeV, we have  $\langle \frac52^{\langle +\rangle}|\pi|\frac52^{\langle +\rangle}\rangle=0.433$ and $\langle \frac32^{\langle +\rangle}|\pi|\frac32^{\langle +\rangle}\rangle=0.241$. In the former case, the largest two Nilsson contributions to the $5/2^{\langle +\rangle}$ state are $5/2^+[633]$ and $5/2^+[642]$, whereas the largest two Nilsson components of the $5/2^{\langle -\rangle}$ are $5/2^-[752]$ and $5/2^-[743]$. The octupole coupling between these two $5/2$ states involves a leading component in one state and a subleading component in the other state. This occurs in two distinct ways, leading to a parity mixing. On the other hand, the dominant Nilsson configuration of the $3/2^{\langle +\rangle}$ state is $3/2^+[651]$ whereas a $3/2^{\langle -\rangle}$ state is found just below $-7.5$~MeV (outside the range plotted in  Fig.~\ref{fig_228Th_spoct}) with a dominant configuration $3/2^-[741]$. The resulting parity mixing between these two $3/2$ states thus occurs essentially through the coupling of these leading Nilsson configurations.

Moreover, the effect of this octupole coupling on the single-particle energies generally preserves the ordering of the states characterized by their average parity, as expected from a reasoning based on second-order perturbation theory. This is typically the case in the pair of $3/2$ states discussed above. Because the coupling between them is strong, the single-particle energy of the $3/2^{\langle +\rangle}$ state significantly increases while that of the $3/2^{\langle - \rangle}$ drops. However, the $5/2^{\langle -\rangle}$ state, expected to lie above the Fermi level and above the $5/2^{\langle +\rangle}$ state for very small octupole moments, happens to drop far below the Fermi level for the actual octupole deformation. At the same time the $3/2^{\langle +\rangle}$ state gets pushed above the $5/2^{\langle +\rangle}$ state, both lying above the Fermi level. By a similar mechanism a $7/2^{\langle -\rangle}$ state is pushed down from far above the Fermi level in the symmetric solution, in such a way that it lies between the $5/2^{\langle +\rangle}$ and $3/2^{\langle +\rangle}$ states just above the Fermi level.

This overall structure of the s.p. spectrum in $^{228}$Th suggests that the  $5/2^{+}$ and  $3/2^{+}$ states may be expected to lie close to each other in the odd-mass nucleus $^{229}$Th. When octupole deformations are taken into account (for the reflection-asymmetric solution), in addition also the $7/2^-$ state is expected to be present in the same vicinity.   Comparing Figs.~\ref{fig_228Th_sprefsym} and \ref{fig_228Th_spoct} we remark that the spacing and order of the s.p. levels are more strongly affected by the inclusion of octupole deformation rather than by the fine (re)adjustment of the pairing strengths $G_n$ and $G_p$.

\section{$^{229}$Th calculation}
\label{th229calc}

\subsection{Calculations of one-quasiparticle states with
  self-consistent blocking}

We perform the HFBCS calculations for $^{229}$Th by separately blocking the orbitals near the $^{228}$Th Fermi level in accordance to the level structures in Figs.~\ref{fig_228Th_sprefsym} and \ref{fig_228Th_spoct}. These are the $5/2^{+}$, $3/2^{+}$, $5/2^{-}$ and $7/2^{-}$ orbitals. The numerical calculations are performed in a regime of broken time-reversal invariance (see Appendix~A for the time-odd terms used) with removed Kramers degeneracy so that each s.p. orbital splits into a pair of positive and negative $K$ counterparts (see Ref.~\cite{PRC15_hfbcs} for more details).

Blocking a single-particle state with a positive $K$ quantum number or its pairing partner with the opposite value of the $K$ quantum number leads to time-reversed solutions with the same expectation values of time-even operators, including the total energy. Therefore one can consider only one-quasiparticle states with positive $K$ values in the blocking procedure.

\begin{table*}[ht]
\begin{center}
\caption{
Total energy, $E_{\mbox{\scriptsize tot}}(K^{\pi})$, excitation
energies, $E^{*}(K^{\pi})$, Eq.~(\ref{Eiso}), neutron ($\Delta_{n}$)
and proton ($\Delta_{p}$) pairing gaps, mass and charge quadrupole
moments ($Q_{20}^m$ and $Q_{20}^p$) and corresponding deformation
parameters $\beta_{2}^{m}$ and $\beta_{2}^{p}$, magnetic moment
($\mu$), collective gyromagnetic factor $g_R$ and spin-gyromagnetic
quenching $q_s$ obtained by reflection symmetric SIII-HFBCS
calculations for blocked $5/2^{+}$, $3/2^{+}$, $5/2^{-}$ and $7/2^{-}$
orbitals in $^{229}$Th with the set of pairing strengths  $G_n=16.0$
MeV and $G_p=14.4$ MeV. For each orbital the Nilsson quantum numbers of the leading component and its contribution in the s.p. wave function (in {\%}) is given  (see text for further details).}
\label{tab:th229reflsym}
\bigskip
\tabcolsep=3pt
\begin{tabular}{ccccccccccccc}
\hline\hline
\noalign{\smallskip}
$K^{\pi}$ &$[Nn_{z}\Lambda]$({\%}) & $E_{\mbox{\scriptsize tot}}$&$E^{*}$ &$\Delta_{n}$&$\Delta_{p}$& $Q_{20}^m$ & $\beta_{2}^{m}$ & $Q_{20}^p$ & $\beta_{2}^{p}$ & $\mu$  & $g_R$ & $q_s$ \\
 & &  (MeV)    &  (keV)   & (MeV)        & (MeV)        &    (fm$^2$)     & &    (fm$^2$)     &  & ($\mu_N$)&       &        \\
\hline
$\frac{5}{2}^{+}$& [633] (56.4)&
$-$1743.3668 & 0.0  & 0.642 & 0.882 & 2150.9 & 0.217 & 850.6 & 0.219 & 0.7366   & 0.291  & 0.738 \\

$\frac{3}{2}^{+}$& [631] (31.9)&
$-$1743.3497  &17.1& 0.645 & 0.884 & 2147.9 & 0.217 & 849.5 & 0.219 & $-$0.3128& 0.352  & 0.765 \\

$\frac{5}{2}^{-}$& [752] (35.3)&
$-$1743.3499 &16.9& 0.644 & 0.884 & 2144.9 & 0.217 & 848.3 & 0.219 & $-$0.2457 & 0.473  & 0.714 \\

$\frac{7}{2}^{-}$& [743] (52.1)&
$-$1742.4774 &889.4& 0.774 & 0.904 & 2135.0 & 0.216 & 843.2 & 0.217 & $-$0.4659 & 0.368  & 0.735 \\

\hline\hline
\end{tabular}
\end{center}
\end{table*}

The excitation energies (q.p. bandheads) are determined as
\begin{eqnarray}
E^{*}(K^{\pi})= E_{\mbox{\scriptsize tot}}(K^{\pi})-E_{\mbox{\scriptsize tot}}(K^{\pi}_{\mbox{\scriptsize GS}}) \ ,
\label{Eiso}
\end{eqnarray}
where $E_{\mbox{\scriptsize tot}}(K^{\pi}_{\mbox{\scriptsize GS}})$ is the lowest total energy corresponding to the GS orbital.

The calculations were made in the cases of imposed and released reflection symmetry. Calculations with selfconsistent blocking and intrinsic parity breaking are very scarce in the literature~\cite{Ban10}. Their convergence is made particularly difficult when two single-particle states with the same $K$ quantum number but opposite parity content are both close to the chemical potential. This precisely happens in the case of the $5/2$ one-quasiparticle states of the $^{229}$Th nucleus. In the same spirit as explained in Ref.~\cite{Afanasjev13} about selfconsistent blocking in Cranked Hartree--Bogoliubov + Lipkin--Nogami calculations, several ``fingerprints'' have to be used to identify the state to be blocked. In the present work we use the average parity $\expval{\pi}$ in addition to the $K$ quantum number. Even though $\expval{\pi}$ turns out to be almost vanishing in some cases, it proves to be efficient to make the blocked calculations converge in the course of Hartree--Fock iterations.

\subsection{Numerical results}

\begin{figure*}[ht]
\centering
\includegraphics[width=16cm]{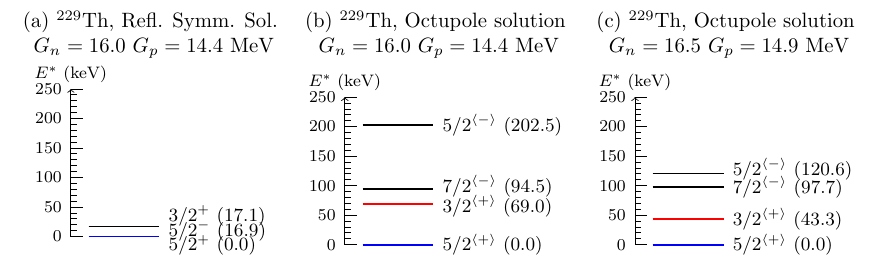}
\caption{Lowest one-neutron q.p. excitation energies [see Eq.~(\ref{Eiso})],
  in $^{229}$Th calculated in (a) reflection-symmetric
  (see also Table~\ref{tab:th229reflsym}) and (b), (c) reflection-unconstrained, asymmetric
  (see also Table~\ref{tab:th229oct}) Skyrme-SIII HFBCS approximation for two
  sets of pairing constants. In panel (a) the lines depicting the
  $3/2^{+}$ and $5/2^{-}$ states are degenerate at the given
  scale and the $7/2^{-}$ level at 889.4 keV (see
  Table~\ref{tab:th229reflsym}) is out of the plot range.}
\label{fig_229Th_excit_levels}
\end{figure*}

The obtained numerical results of the reflection-symmetric and asymmetric calculations are given in Tables~\ref{tab:th229reflsym} and \ref{tab:th229oct}, respectively. The corresponding one-neutron q.p. excitation (bandhead) energy levels are illustrated in Fig.~\ref{fig_229Th_excit_levels}. In both cases the obtained $5/2^{+}$ state has the lowest energy and thus correctly represents the experimentally observed $^{229}$Th ground state.

Our results for the reflection symmetric case are presented in Table~\ref{tab:th229reflsym} and Fig.~\ref{fig_229Th_excit_levels}(a). Summarizing our findings, we note that:

(i) The $E^{*}(3/2^{+})$ excitation energy corresponding to the $^{229m}$Th isomer appears at $17.1$ keV. Although it is 3 orders of magnitude larger than the experimental isomer energy of 8 eV, from nuclear structure point of view the model $3/2^{+}$ state appears almost degenerate with the $5/2^{+}$ ground state.  Note that the SIII effective interaction used reproduces the Nilsson quantum numbers of the leading components in the s.p. wave function of the two orbitals 5/2[633] and 3/2[631] (see Table~\ref{tab:th229reflsym}) widely adopted in the literature and experimental data bases \cite{ensdf}.

(ii) Regarding the other two considered orbitals, we see that while the $E^{*}(7/2^{-})$ energy is about 0.9 MeV, obviously out of the range of very low-energy excitations, the energy of the $5/2^{-}$ state is almost degenerate with and slightly below the $3/2^{+}$ isomer. From the model point of view this can be understood by the almost degenerate $5/2^{+}$ and $5/2^{-}$ orbitals observed in Fig.~\ref{fig_228Th_sprefsym} for $^{228}$Th. On the other hand the obtained $5/2^{-}$ q.p. excitation in $^{229}$Th can be associated with the $5/2[752]$ bandhead, identified in the experiment with energy of about 143 keV \cite{ensdf}. The latter is considered as the negative-parity counterpart of the 5/2[633] ground state within the yrast quasi-parity-doublet of $^{229}$Th. Here the energy of this level is essentially underestimated, but as shown later on, this result considerably changes with the inclusion of the octupole deformation.

(iii) The calculations also provide predictions for the quadrupole moments and deformation, the magnetic dipole moments $\mu$, the collective gyromagnetic factor of the even-even core $g_R$ and the spin-gyromagnetic quenching $q_s$ in the four considered q.p. states of $^{229}$Th. Details on the calculation of the magnetic moments and gyromagnetic factors with the quenching effect are given in Secs. IIC, IIID and Appendix B of Ref.~\cite{PRC15_hfbcs}.  From Table~\ref{tab:th229reflsym} we see that, similarly to the DSM-CQOM calculations \cite{Minkov_Palffy_PRL_2019,Minkov_Palffy_PRC_2021}, the obtained GS magnetic moment of about $0.74\mu_{N}$ overestimates the experimental values of $0.360(7)\mu_N$ \cite{Safronova13} and $\mu_{\mbox{\scriptsize GS}}=0.45(4)\mu_N$ \cite{Gerstenkorn74}. On the other hand, again similarly to DSM-CQOM, the obtained IS magnetic moment of $-0.31$ $\mu_N$ agrees well with the recent experimental value of $-0.37(6)\mu_N$ \cite{Thielking2018,Mueller18}.

We note that the present HFBCS calculations provide different collective and spin gyromagnetic factors for the different blocked orbitals. The collective gyromagnetic factor $g_R$ is calculated for the polarized even-even core in the cranking approximation with BCS pairing excluding the blocked state and its conjugate state (for the definition of the latter see Sec. IIA of Ref.~\cite{PRC15_hfbcs}). On the other hand the DSM-CQOM calculations \cite{Minkov_Palffy_PRL_2019,Minkov_Palffy_PRC_2021} use common gyromagnetic factors for the ground and isomeric state taken as phenomenological parameters. Therefore, only a rough comparison of the magnetic moments and gyromagnetic factors in both HFBCS and DSM-CQOM approaches can be made.

\begin{table*}[ht]
\begin{center}
\caption{
The same as in Table~\ref{tab:th229reflsym}, and in addition the mass octupole moment $Q_{30}^{m}$ and deformation
$\beta_{3}^{m}$. All quantities were obtained by reflection-unconstrained (asymmetric) HFBCS
calculations for blocked $5/2^{\langle +\rangle}$, $3/2^{\langle
  +\rangle}$, $5/2^{\langle -\rangle}$ and $7/2^{\langle -\rangle}$
orbitals in $^{229}$Th with the two different sets of pairing
strengths, Set 1 ($G_n=16.0$ MeV, $G_p=14.4$ MeV) and Set 2
($G_n=16.5$ MeV, $G_p=14.9$ MeV). The solution for $5/2^{\langle
  -\rangle}$ state, obtained with Set 2, corresponds to zero octupole
deformation. For each blocked orbital the average parity $\langle\pi\rangle$ is also given. See text for further explanations.}
\label{tab:th229oct}
\bigskip
\tabcolsep=2pt
{\small
\begin{tabular}{cccrccccccccccccc}
\hline\hline
\noalign{\smallskip}
$K^{\pi}$ & $G_p,G_n$ & $[Nn_{z}\Lambda]$({\%})&$\langle\pi\rangle$&$E_{\mbox{\scriptsize tot}}$&$E^{*}$&$\Delta_{n}$&$\Delta_{p}$&$Q_{20}^m$&$\beta_{2}^{m}$ & $Q_{20}^p$&$\beta_{2}^{p}$ &$Q_{30}^{m}$ & $\beta_{3}^{m}$& $\mu$  & $g_R$ & $q_s$ \\
 & Set &   & & (MeV)    &  (keV)   & (MeV)        & (MeV)        & (fm$^2$)   &  &   (fm$^2$)   &  & ($b^{3/2}$)& & ($\mu_N$)&       &        \\
\hline
\multirow{2}{*}{$\frac{5}{2}^{\langle +\rangle}$}&
Set 1 &[633] (43.3)&0.350& $-$1743.5399 & 0.0  & 0.542 & 0.814 & 2107.3 & 0.213 &832.0 & 0.214 & 2.369 & 0.110 & 0.4254& 0.281 & 0.788\\

&Set 2&[633] (41.7)&0.297& $-$1743.8174 & 0.0  & 0.689 & 0.933 & 2107.0 & 0.213 &831.9 & 0.214 & 2.141 & 0.099 & 0.4181& 0.303 & 0.816
\vspace{0.2cm}\\

\multirow{2}{*}{$\frac{3}{2}^{\langle +\rangle}$}&
Set 1 &[631] (22.6)&0.530& $-$1743.4709 &69.0& 0.568 & 0.851 & 2136.8 & 0.216 &843.7 & 0.217 & 1.558 & 0.072 & $-$0.2729 & 0.256  & 0.677 \\

&Set 2&[631] (23.8)&0.593& $-$1743.7741 & 43.3 & 0.705 & 0.963 & 2133.6 & 0.216 &842.5 & 0.217 & 1.334 & 0.062 & $-$0.2635& 0.316 & 0.673
\vspace{0.2cm}\\

\multirow{2}{*}{$\frac{5}{2}^{\langle -\rangle}$}&
Set 1& [752] (23.6)&$-$0.330&$-$1743.3379 & 202.5 & 0.644 & 0.885 & 2144.4 & 0.217 &848.0 & 0.219 & 0.182 & 0.008 & 0.0408& 0.405 & 0.825\\

&Set 2& [752] (35.3)&$-$1.000&$-$1743.6968 & 120.6 & 0.744 & 0.986 & 2140.7 & 0.216 &846.2 & 0.218 & 0.000 & 0.000 & $-$0.2551& 0.457 & 0.716
\vspace{0.2cm}\\

\multirow{2}{*}{$\frac{7}{2}^{\langle -\rangle}$}&
Set 1 &[743] (32.5)&$-$0.370& $-$1743.4454 & 94.5 & 0.581 & 0.768 & 2062.0 & 0.208 &815.3 & 0.210 & 3.122 & 0.144 & $-$0.1018 & 0.400  & 0.725 \\

&Set 2&[743] (32.7)&$-$0.374& $-$1743.7197 & 97.7 & 0.714 & 0.881 & 2049.6 & 0.207 &810.4 & 0.209 & 3.004 & 0.139 & $-$0.0880& 0.410 & 0.719
\vspace{0.2cm}\\

\hline\hline
\end{tabular}
}
\end{center}
\end{table*}

The results of our reflection-unconstrained calculations are presented in Table~\ref{tab:th229oct} and illustrated in Figs.~\ref{fig_229Th_excit_levels}(b) and (c). Summarizing our findings for this part, we note that:

(i) Comparing the total energies obtained for pairing strengths Set 1 ($G_n=16.0$ MeV, $G_p=14.4$ MeV) in Tables \ref{tab:th229reflsym} and \ref{tab:th229oct} we notice that the octupole solutions for the $5/2^{\langle +\rangle}$, $3/2^{\langle +\rangle}$, $5/2^{\langle -\rangle}$ and $7/2^{\langle -\rangle}$ states in Table~\ref{tab:th229oct} appear deeper in energy by 173, 121, 12 and 968 keV, respectively, compared to those in the reflection-symmetric case of Table~\ref{tab:th229reflsym}.  This result favours the presence of octupole deformation in the $5/2^{\langle +\rangle}$, $3/2^{\langle +\rangle}$,  and $7/2^{\langle -\rangle}$ states, for which the model predicts non-zero octupole moments. For Set 2 of $G_n=16.5$ MeV, $G_p=14.9$ MeV we observe additional deepening of the total energy which is essentially due to the change in the pairing energy contribution. The $5/2^{\langle -\rangle}$ state presents the smallest octupole deformation for Set 1 and even zero octupole deformation for Set 2.

(ii) From Table~\ref{tab:th229oct} we notice that the three blocked orbitals appear with mixed average parity. In the $5/2^{\langle +\rangle}$ GS the octupole deformation of $\beta_{3}\approx 0.10-0.11$ is obtained a bit larger than that in the $3/2^{\langle +\rangle}$ IS where $\beta_{3}\approx 0.06-0.07$.  We note that in the DSM-CQOM approach \cite{Minkov_Palffy_PRL_2017} both states are considered with the same quadrupole and octupole deformations $\beta_{2}=0.240$ and $\beta_{3}=0.115$. Comparing the results of the two approaches we notice that the quadrupole deformations obtained for the different states in Table~\ref{tab:th229oct} are a bit smaller than that in DSM-CQOM whereas the octupole deformation in the $5/2^{\langle +\rangle}$ GS appears quite close to that in DSM-CQOM.

(iii) Regarding the excited level spacings, we see in Figs.~\ref{fig_229Th_excit_levels}(b) and (c) (see also Table~\ref{tab:th229oct}) that the $3/2^{\langle + \rangle}$ state appears as the lowest excitation corresponding to the isomer with an energy of $69$ keV for the Set 1 ($G_n=16.0$ MeV, $G_p=14.4$ MeV) and $43$ keV for Set 2 ($G_n=16.5$ MeV, $G_p=14.9$ MeV). Both values are larger than that in the reflection-symmetric case in Fig.~\ref{fig_229Th_excit_levels}(a) (see also Table \ref{tab:th229oct}), but again, from nuclear structure point of view they correspond to a very-low-energy one-q.p. excitation. However, we note that unlike $^{228}$Th, in $^{229}$Th the fine variation in the pairing constants leads to sizable displacements of the s.p. orbitals resulting in a decrease of about 20 keV in the $3/2^{+}$ excitation energy. This increased sensitivity of the q.p. spectrum to the pairing strengths (compared to the even-even core nucleus) might be related to the effect of self-consistent blocking.
About the $5/2^{\langle - \rangle}$ excitation we notice that now it appears much higher than the $3/2^{\langle + \rangle}$ state in contrast to the reflection-symmetric case in which the $5/2^{-}$ state is calculated to be a bit below the $3/2^{+}$ state [Fig.~\ref{fig_229Th_excit_levels} (a)]. We note that the theoretical values for the $5/2^{\langle - \rangle}$ energy given in Table~\ref{tab:th229oct} for the two $(G_n,G_p)$ sets, 121 keV and 203 keV, now bracket its experimental value of 143 keV \cite{ensdf}. The $7/2^{\langle - \rangle}$ excitation now appears lower, around 90--100 keV, compared to its much higher energy of 889.4 keV obtained in the reflection-symmetric case (Table~\ref{tab:th229reflsym}). The above results illustrate the importance of both the shape deformation and pairing correlations for the accuracy of the model predictions and the need of their finer treatment for the further quantitative description of $^{229m}$Th properties.

(iv) Regarding the GS and IS magnetic moments, we notice that now the obtained GS value $\mu_{\mbox{\scriptsize GS}}\approx 0.42-0.43$ $\mu_N$ appears lower compared to the reflection-symmetric case, already stepping between the two available (new and old) experimental values of $0.360(7)$ $\mu_N$ \cite{Safronova13} and 0.45(4) $\mu_N$ \cite{Gerstenkorn74}. Furthermore, this is the lowest value reached in the DSM-CQOM approach (see Fig.~10(d) in Ref.~\cite{Minkov_Palffy_PRC_2021}). At the same time, the calculated isomeric magnetic moment $\mu_{\mbox{\scriptsize IS}}$ of about $-$0.27 $\mu_N$ underestimates in absolute value the corresponding experimental value of $-0.37(6)\mu_N$ \cite{Thielking2018,Mueller18}. These results look consistent with the DSM-CQOM calculations \cite{Minkov_Palffy_PRC_2021}. We note, however, that the corresponding spin-gyromagnetic quenching $q_s$, varying between 0.7 and 0.8 (see Table~\ref{tab:th229oct}), is larger compared to the value of 0.6 considered in \cite{Minkov_Palffy_PRL_2019,Minkov_Palffy_PRC_2021}. Furthermore, comparing the collective gyromagnetic factor $g_R\approx 0.28 - 0.30$ obtained for $5/2^{\langle + \rangle}$ with the phenomenological $^{229}$Th value of $g_R^{\mbox{\scriptsize phen}}=Z/(Z+N)=0.393$ we observe a quenching of the latter by a factor of $q_R=g_R/g_R^{\mbox{\scriptsize phen}}\approx 0.71 - 0.76$. For the $3/2^{\langle + \rangle}$ isomer state this quenching factor is $q_R\approx 0.65 - 0.80$. These values are larger than the corresponding ones considered in CQOM-DSM \cite{Minkov_Palffy_PRL_2019,Minkov_Palffy_PRC_2021}. Thus, it seems that the present calculations are able to approach the corresponding experimental data (especially for the GS) through smaller $g_R$ attenuation (larger $q_R$). These findings highlight an advantage of the microscopic HFBCS approach over the more phenomenological CQOM-DSM one, since there is no need for phenomenological quenching factors, and the theoretical predictions are close to the experimental values already at smaller attenuation. We note that a more accurate comparison between the two CQOM-DSM and HFBCS approaches can be made on this point after a parity projection is implemented in the latter, although it has been shown that projection after variation affects the total energy only slightly \cite{Nhan2012}.

Concluding on the results obtained so far, the present microscopic approach implemented without particular effort to achieve realistic $^{229m}$Th description, shows that the applied HFBCS scheme may carry the basic features needed to study this isomer from a deeper nuclear many-body structure perspective. We emphasize that the quadrupole and octupole deformations $\beta_2$ and $\beta_3$ are not free parameters in the EDF approach, but appear as an output of the self-consistent variation procedure. On the other hand, the results obtained suggest that their constraining together with tuning of the pairing strengths at further stage may allow us to determine the proximity of the $5/2^{\langle +\rangle}$ and $3/2^{\langle +\rangle}$ orbitals so as to approach the $^{229m}$Th energy scale. At the same time the reasonable values obtained for the magnetic moments show that in a further perspective the model could be able to provide reliable predictions for the $B(M1)$ isomer decay rates. However, the more realistic description of the $^{229m}$Th properties within the HFBCS approach requires further development involving Coriolis mixing and appropriate coupling to the collective degrees of freedom as, for example, done in the CQOM-DSM scheme.

\section{Calculations in neighboring nuclei}
\label{neighb}

\begin{table*}
\begin{center}
\caption{
The same as in Tables~\ref{tab:th229reflsym} and \ref{tab:th229oct}
but for the reflection-symmetric and reflection-unconstrained
(octupole) HFBCS solutions for the lowest (below 1 MeV) q.p. states in
$^{227}$Th, $^{231}$Th, $^{227}$Ra and $^{231}$U. The symbol
``$\dag$'' marks results of reflection-unconstrained calculations
reaching solutions with reflection symmetry.}
\label{tab:neighbors}
\bigskip
\tabcolsep=2pt
{\small
\begin{tabular}{cccccccccccccc}
\hline\hline
\noalign{\smallskip}
$K^{\pi}$& $E_{\mbox{\scriptsize tot}}$&$E^{*}$&$\Delta_{n}$&$\Delta_{p}$&$Q_{20}^m$&$\beta_{2}^{m}$ & $Q_{20}^p$&$\beta_{2}^{p}$ &$Q_{30}^{m}$ & $\beta_{3}^{m}$& $\mu$  & $g_R$ & $q_s$ \\
  &  MeV    &  keV   & MeV        & MeV        & fm$^2$   &  &   fm$^2$   &  & $b^{3/2}$& & $\mu_N$&       &        \\
\hline

\multicolumn{14}{c}{$^{227}$Th, \ $G_n=16.5$ MeV, $G_p=14.9$ MeV, \ $b=0.480$, $q=1.20$} \\

$\frac{1}{2}^{+}$&
$-$1730.0238 & 941.7 & 0.934 & 1.019 & 1816.5 & 0.187 &729.6 & 0.190 & -- & -- & $-$0.1017& 0.334 & 0.763\\

$\frac{3}{2}^{+}$&
$-$1730.9523 & 13.2 & 0.749 & 0.977 & 1961.4 & 0.202 &786.3 & 0.204 & -- & -- & $-$0.2968& 0.339 & 0.766\\

$\frac{3}{2}^{-}$&
$-$1730.8456 & 119.9 & 0.839 & 1.003 & 1865.3 & 0.192 &748.5 & 0.194 & -- & -- & $-$0.1110& 0.400 & 0.725\\

$\frac{5}{2}^{+}$&
$-$1730.9655 & 0.0 & 0.769 & 0.988 & 1943.2 & 0.200 &779.4 & 0.202 & -- & -- & $-$0.7080& 0.272 & 0.741\\

$\frac{5}{2}^{-}$&
$-$1730.8853 & 80.2 & 0.760 & 0.973 & 2002.2 & 0.206 &802.5 & 0.208 & -- & -- & $-$0.2568& 0.444 & 0.708

\vspace{0.1cm}\\

$\frac{1}{2}^{\langle +\rangle}$&
$-$1731.7831 & 0.0 & 0.571 & 0.835 & 1877.0 & 0.192 &754.9 & 0.196 & 3.044 & 0.143 & $-$0.0860& 0.343 & 0.718\\
$\frac{3}{2}^{\langle +\rangle}$&
$-$1731.4371 & 346.0 & 0.826 & 0.842 & 1803.8 & 0.185 &726.3 & 0.188 & 3.023 & 0.142 & 0.4098& 0.267 & 0.737\\

$\frac{5}{2}^{\langle +\rangle}$&
$-$1731.3871 & 396.0 & 0.737 & 0.853 & 1794.1 & 0.184 &722.7 & 0.187 & 2.849 & 0.134 & 0.4665& 0.396 & 0.254\\

$\frac{7}{2}^{\langle -\rangle}$&
$-$1731.4324 & 350.7 & 0.677 & 0.836 & 1785.0 & 0.183 &721.5 & 0.188 & 3.282 & 0.154 & 0.0129& 0.454 & 0.755

\vspace{0.2cm}\\

\multicolumn{14}{c}{$^{231}$Th, \ $G_n=15.8$ MeV, $G_p=14.3$ MeV, \ $b=0.480$, $q=1.20$} \\

$\frac{3}{2}^{+}$&
$-$1755.3403 & 8.6 &  0.504 & 0.892 & 2308.2 & 0.230 &901.6 & 0.231 & -- & -- & $-$0.3864& 0.268 & 0.761\\

$\frac{5}{2}^{+}$&
$-$1755.3489 & 0.0 & 0.483 & 0.875 & 2327.3 & 0.231 &907.9 & 0.233 & -- & -- & $-$0.6816& 0.186 & 0.734\\

$\frac{5}{2}^{-}$&
$-$1755.3467 & 2.2 & 0.507 & 0.894 & 2260.0 & 0.225 &881.9 & 0.226 & -- & -- & $-$0.3160& 0.377 & 0.721

\vspace{0.1cm}\\

$\frac{3}{2}^{\langle +\rangle}$&
$-$1755.3746 & 0.0 & 0.536 & 0.884 & 2294.2 & 0.228 &895.2 & 0.229 & 1.038 & 0.047 & $-$0.3414& 0.291 & 0.717\\

$\frac{5}{2}^{\langle +\rangle \dag}$&
$-$1755.3490 & 25.6 & 0.483 & 0.875 & 2327.3 & 0.231 &907.9 & 0.233 & 0.000 & 0.000 & $-$0.6816& 0.186 & 0.734\\

$\frac{5}{2}^{\langle -\rangle \dag}$&
$-$1755.3467 & 27.9 & 0.507 & 0.894 & 2260.0 & 0.225 &881.9 & 0.226 & 0.000 & 0.000 & $-$0.3160& 0.377 & 0.721

\vspace{0.2cm}\\

\multicolumn{14}{c}{$^{227}$Ra, \ $G_n=15.9$ MeV, $G_p=14.4$ MeV, \ $b=0.480$, $q=1.15$}\\

$\frac{5}{2}^{+}$& $-1731.507$& 0  & 0.670 & 0.893 & 1914.2 & 0.197 &728.7 & 0.195 & -- & -- & 0.6399& 0.202 & 0.747 \\

$\frac{3}{2}^{+}$& $-1731.472$ &35& 0.675 & 0.894 & 1910.1 & 0.196 &727.4 & 0.194 & -- & -- & $-0.3259$ & 0.271  & 0.760 \\

$\frac{5}{2}^{-}$& $-1731.458$& 49 & 0.668 & 0.894 & 1917.3 & 0.197 &730.0 & 0.195 & -- & -- &$-$0.2871& 0.291 & 0.710

\vspace{0.1cm}\\

$\frac{5}{2}^{\langle +\rangle}$& $-1731.535$ & 36 & 0.629 & 0.804 & 1921.0 & 0.197 &733.5 & 0.196 & 1.699 & 0.080 & 0.3504& 0.298 & 0.947 \\

$\frac{3}{2}^{\langle +\rangle}$& $-1731.571$ & 0 & 0.624 & 0.824 & 1924.7 & 0.198 &733.6 & 0.196 & 1.287 & 0.061 & $-0.2484$ & 0.210  & 0.670 \\

$\frac{7}{2}^{\langle -\rangle}$& $-1731.337$ & 234 & 0.652 & 0.774 & 1884.9 & 0.193 &723.1 & 0.193 & 2.585 & 0.122 & $-0.1181$ & 0.388  & 0.734 

\vspace{0.2cm}\\

\multicolumn{14}{c}{$^{231}$U, \ $G_n=15.9$ MeV, $G_p=14.4$ MeV, \ $b=0.480$, $q=1.20$}\\

$\frac{5}{2}^{-}$& $-1753.612$& 27 & 0.639 & 0.823 & 2321.0 & 0.231 &945.3 & 0.236 & -- & -- & $-0.2091$ & 0.544& 0.717\\

$\frac{5}{2}^{+}$& $-1753.639$& 0  & 0.637 & 0.817 & 2339.5 & 0.232 &952.4 & 0.237 & -- & -- & 0.8170& 0.362 & 0.731 \\

$\frac{7}{2}^{-}$& $-1752.790$&849 & 0.782 & 0.840 & 2359.2 & 0.234 &957.8 & 0.239 & -- & -- & $-0.4024$ & 0.470& 0.736 \\

$\frac{3}{2}^{+}$& $-1753.620$& 19 & 0.640 & 0.820 & 2336.1 & 0.232 &951.1 & 0.237 & -- & -- & $-0.3009$ & 0.420& 0.766 

\vspace{0.1cm}\\

$\frac{5}{2}^{\langle -\rangle \dag}$& $-1753.612$& 70 & 0.639 & 0.823 & 2321.0 & 0.231 &945.3 & 0.236 & 0.000 & 0.000 & $-0.2091$ & 0.544 & 0.717\\

$\frac{5}{2}^{\langle +\rangle}$&
$-1753.682$ & 0 & 0.503 & 0.804 & 2288.6 & 0.227 &930.6 & 0.232 & 2.205 & 0.100 & $0.4623$ & 0.319  & 0.762 \\

$\frac{7}{2}^{\langle -\rangle}$&
$-1753.559$ & 123 & 0.517 & 0.714 & 2222.2 & 0.220 &902.5 & 0.225 & 3.321 & 0.151 & $-0.0640$ & 0.441 & 0.692 \\

$\frac{3}{2}^{\langle +\rangle}$&
$-1753.649$ & 33 & 0.601 & 0.825 & 2326.3 & 0.231 &946.7 & 0.236 & 1.065 & 0.048 & $-0.2545$ & 0.396  & 0.676 \\

\hline\hline
\end{tabular}
}
\end{center}
\end{table*}

Our results show that the low-energy isomer properties of $^{229}$Th are determined by the position of the $5/2^{+}$ and $3/2^{+}$ neutron orbitals near the Fermi level. Thus,  it is worth examining the evolution of this condition in the neighboring isotopes, $^{227}$Th and $^{231}$Th, as well as in the isotones $^{227}$Ra and $^{231}$U. For each nucleus we calculate the $5/2^{+}$ and  $3/2^{+}$ neutron q.p. excitation energies together with few neighboring low-lying one-neutron excitations. The basis parameters and pairing constants are determined for the even-even cores ($^{226}$Th, $^{230}$Th, $^{226}$Ra and $^{230}$U) by keeping the theoretical moment of inertia [respectively the $E(2_{1}^{+})$ energies] close to the experimental values, following the prescription applied for $^{229}$Th. The numerical results obtained by the reflection-symmetric and reflection-unconstrained HFBCS calculations are collected in Table~\ref{tab:neighbors}. Their detailed analysis is given in the following.

\begin{figure*}[ht]
\centering
\includegraphics[width=10cm]{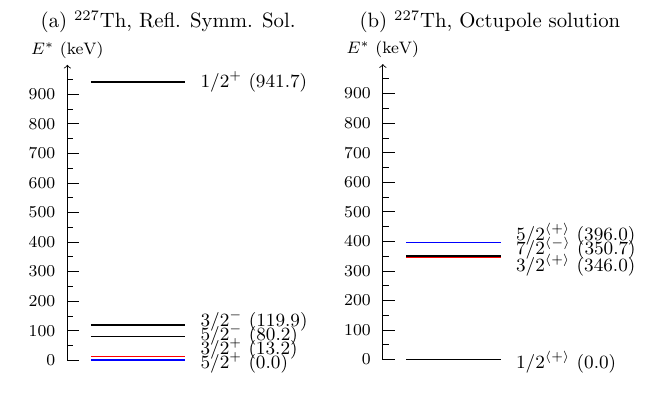}\\
\includegraphics[width=10cm]{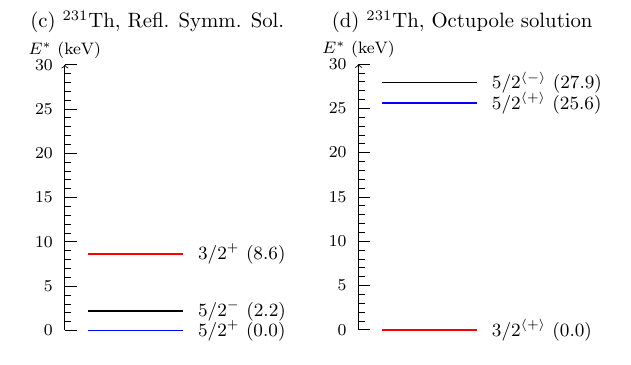}
\caption{Lowest one-neutron q.p. excitation energies [given by
  Eq.~(\ref{Eiso})] in $^{227}$Th and $^{231}$Th calculated in the
  (a), (c) reflection-symmetric and (b), (d) reflection-unconstrained
  Skyrme-SIII HFBCS approximation, respectively (see also
  Table~\ref{tab:neighbors}). The indistinguishable level bars are
  identified by the spin labels.}
\label{fig_Th227_231}
\end{figure*}

\subsection{Neighboring isotopes: $^{227}$Th and $^{231}$Th}

In $^{227}$Th we found solutions below 1 MeV for the $5/2^{+}$ and $3/2^{+}$ orbitals of interest and for the $1/2^{+}$ orbital. In addition, we find below 1 MeV the $3/2^{-}$ and $5/2^{-}$ orbitals only in the reflection-symmetric solution, and the $7/2^{\langle - \rangle}$ orbital only in the octupole solution, respectively, as illustrated in Figs.~\ref{fig_Th227_231}(a)-(b). We observe that the reflection-unconstrained solutions appear strongly favored in energy, with the $1/2^{\langle +\rangle}$, $3/2^{\langle +\rangle}$ and $5/2^{\langle +\rangle}$ states lowered by about 1.8 MeV, 0.5 MeV and 0.4 MeV, respectively, with respect to the reflection-symmetric ones (see Table~\ref{tab:neighbors}). As a result the octupole $1/2^{\langle +\rangle}$ state becomes the ground state, instead of the $5/2^{\langle +\rangle}$ state in $^{229}$Th, in agreement with the experimental data \cite{ensdf}. In all considered states the reflection-asymmetric solution gives large octupole moments and deformations, $\beta_{3}\approx 0.13-0.15$, suggesting firm octupole deformed shapes in the corresponding excitations. The two states $5/2^{+}$ and $3/2^{+}$ remain close to one another by only exchanging their positions between the reflection-symmetric and octupole solutions (see Fig.~\ref{fig_Th227_231}).

In $^{231}$Th we find solutions for $3/2^{+}$, $5/2^{+}$ and $5/2^{-}$
neutron q.p. states. In the reflection-unconstrained solutions we
observe quite weak signs of octupole deformation. For the
$3/2^{\langle +\rangle}$ state we have an energy gain of only 34 keV
with a small octupole deformation of $\beta_{3}=0.047$. For
$5/2^{\langle +\rangle}$ and $5/2^{\langle -\rangle}$ the
reflection-unconstrained solutions appear with zero octupole
deformation and coincide with the corresponding reflection-symmetric
$5/2^{+}$ and $5/2^{-}$ solutions (see Table~\ref{tab:neighbors}). We
note that in the set of reflection-unconstrained results the octupole
$3/2^{\langle +\rangle}$ state appears as the GS whereas the experimental data suggests $5/2^{+}$ as GS
\cite{ensdf}. Similarly to $^{227}$Th the two states remain close to
one another [see  Figs.~\ref{fig_Th227_231}(c)-(d)].

The well pronounced octupole deformation in the $^{227}$Th excitations
and the very weak one in $^{231}$Th are consistent with the
corresponding strong and weak octupole deformations we have observed in the
calculations for the even-even core nuclei $^{226}$Th and
$^{230}$Th. In turn, this result is consistent with the earlier
findings that $^{226}$Th belongs to the region of octupole deformed
nuclei around the border of octupole deformation, whereas $^{230}$Th is
situated beyond this border, where essentially soft octupole vibrations can be
observed \cite{Bonat2005}. In addition, our calculations in the
previous section suggest that $^{228}$Th, the core system of the
$^{229m}$Th isomer of interest, is the last nucleus towards the
$N=82-126$ midshell in the series of even-even Th isotopes possessing
stable octupole deformation in the ground state. This is consistent
with the analysis made in Ref.~\cite{NazOland84} (see Fig. 19
therein). Also,  the overall result above suggests that $^{229}$Th may
be situated on the border of octupole deformation in the series of
odd-mass Thorium isotopes.

\begin{figure*}[ht]
\centering
\includegraphics[width=10cm]{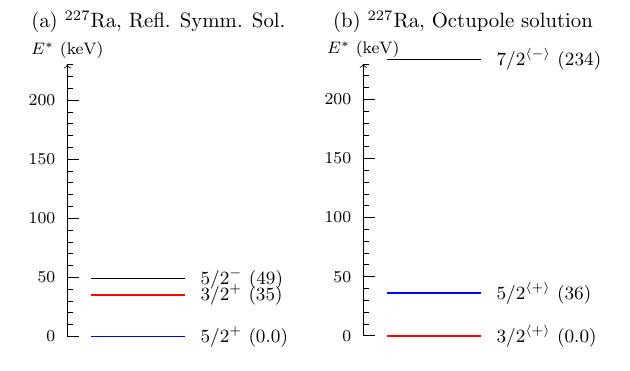}\\
\includegraphics[width=10cm]{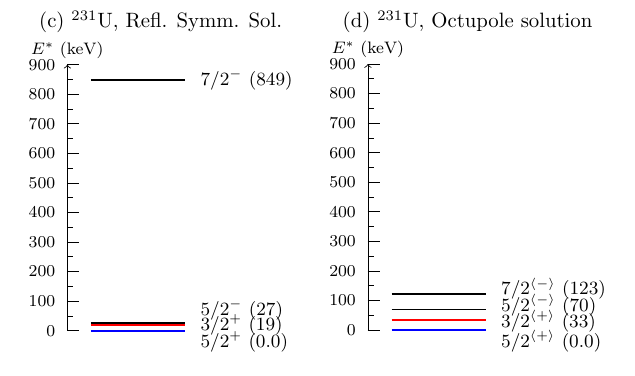}
\caption{Same as in Fig.~\ref{fig_Th227_231}, but for $^{227}$Ra and
  $^{231}$U (see also Table~\ref{tab:neighbors}).}
\label{fig_Ra227_U231}
\end{figure*}

\subsection{Neighboring isotones: $^{227}$Ra and $^{231}$U}

The study of $^{227}$Ra and $^{231}$U, the $N=139$ isotone neighbours
of $^{229}$Th, is of a special interest due to the same number of
neutron orbitals. The core system of the former,
$^{226}$Ra, is also considered to lie around the above mentioned
octupole deformation border and one may expect the core of the latter,
$^{230}$U, to have similar reflection-asymmetric
characteristics. For $^{227}$Ra with experimental GS based on a
$K^{\pi}=3/2^{+}$ s.p. orbital (the same which determines the
$^{229m}$Th IS), a very low lying s.p. excitation of 1.7 keV based on
the $K^{\pi}=5/2^{+}$ orbital (the same which determines the GS of
$^{229}$Th) is reported in \cite{ensdf}. We may therefore
expect that the same microscopic mechanism of quasidegeneracy or
crossing s.p. orbitals identified in $^{229}$Th will play a role for $^{227}$Ra.

In $^{231}$U the experimental situation given in \cite{ensdf} is highly unclear, with the GS supposed to be $5/2^{-}$, while a $5/2^{+}$ state is expected to appear around 40 keV with an uncertainty of 40 keV. No state with $K^{\pi}=3/2^{+}$ is reported in \cite{ensdf} for this nucleus whereas a 45.1(3) keV state with tentative spin $7/2^{-}$ is suggested. On the other hand a very recent measurement, related to population of a rotation band in $^{231}$U \cite{Roux_EPJA2024}, suggests that the ground state is based on the neutron 5/2$^{+}$[633] orbital. Excited bandheads based on the neutron 5/2$^{-}$[752] and 3/2$^{+}$[631] orbitals are also suggested there. These spin values and Nilsson quantum numbers coincide with those appearing in the present analysis.

We are therefore confident that the HFBCS calculations in $^{227}$Ra and $^{231}$U could provide a useful hint on the migration of the $K^{\pi}=5/2^{+}$ and $3/2^{+}$ orbitals and their close neighbors aside of $^{229}$Th
and well as on the evolution of octupole deformation.

\begin{figure*}[ht]
\centering
\includegraphics[width=18cm]{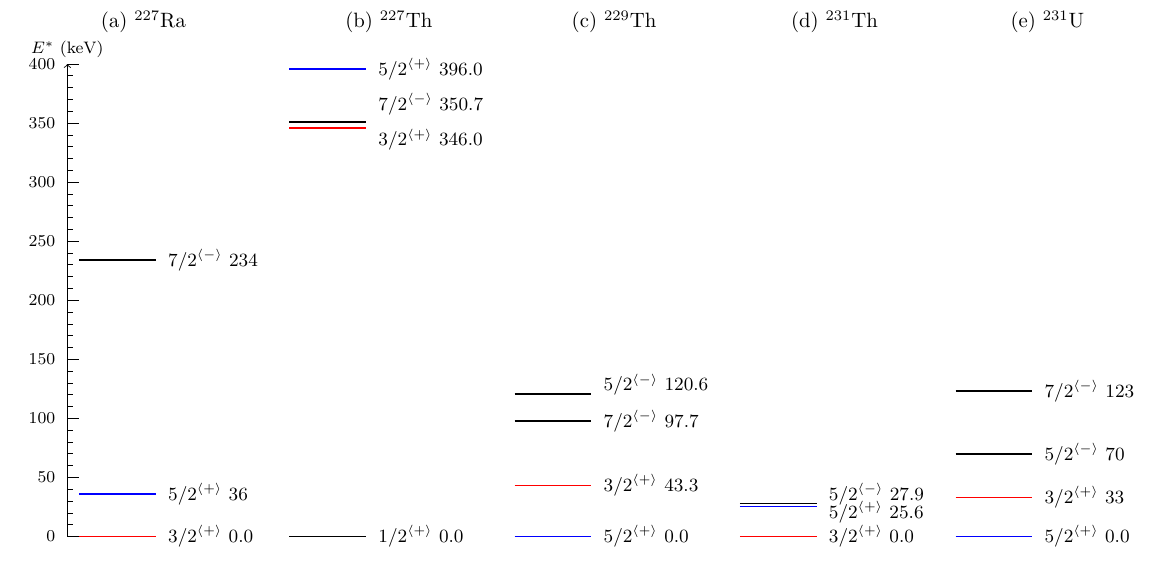}
\caption{Lowest one-neutron q.p. excitation energies [given in
  Eq.~(\ref{Eiso})] in $^{227}$Th, $^{229}$Th, $^{231}$Th, $^{227}$Ra
  and $^{231}$U calculated in the reflection-unconstrained Skyrme-SIII
  HFBCS approximation are given to follow the evolution of the
  $K^{\pi}=5/2^{+}$ and $K^{\pi}=3/2^{+}$ states around  $^{229}$Th
  (see also Table~\ref{tab:neighbors}).}
\label{fig_allnuc}
\end{figure*}

In $^{227}$Ra the reflection-symmetric calculations give the $5/2^{+}$, $3/2^{+}$ and $5/2^{-}$ states with the lowest total energies (below 1 MeV) whereas in the reflection-unconstrained  calculations $5/2^{-}$ is replaced by the $7/2^{\langle -\rangle}$ state. This is illustrated in Fig.~\ref{fig_Ra227_U231}(a)-(b) (see also Table~\ref{tab:neighbors}). We can make the following observations:

(i) For the $5/2^{\langle +\rangle}$ state the reflection-unconstrained (octupole) solution provides an energy gain of only 28 keV compared to the reflection-symmetric solution, with $\beta_{3} = 0.08$. For the $3/2^{\langle +\rangle}$ state the energy gain of the octupole solution is 99 keV with $\beta_{3} = 0.06$. For the $7/2^{-}$ state the calculations suggest some more pronounced octupole deformation with $\beta_3 =0.12$. Nevertheless, the overall result of the SIII HFBCS calculation suggests for $^{227}$Ra a weaker octupole deformation compared to $^{227}$Th. Also it is weaker than what would be expected, given that in the $^{226}$Ra core calculation we got a rather firm octupole deformation. (Note that the latter is considered as an example of a nucleus with a stable octupole deformation in the ground state even though situated around the border of octupole collectivity \cite{BN96,Butler2016}.)  This might be a hint for a stronger influence of the odd neutron under the condition of a different proton number (compared to $^{227}$Th) on the even-even core leading to a softer octupole deformation. More detailed studies (beyond the scope of this work) of the s.p. structure provided by the mean field in this region may be necessary to understand this result.

(ii) Despite the above discussion concerning the octupole deformation, the calculations show that in the reflection-symmetric case the $5/2^{+}$ state appears to be the GS, whereas in the octupole case the GS is given by the $3/2^{\langle +\rangle}$ state (see Fig.~\ref{fig_Ra227_U231}). The latter is favored, even though just slightly, by the values of the total energy and in addition by the assignment in Ref.~\cite{ensdf} which suggests the $^{227}$Ra GS with $K^{\pi}=3/2^{+}$. Also, the octupole calculation favors $5/2^{\langle +\rangle}$ as the first excited q.p. state, again in corroboration with the experiment \cite{ensdf}. We note that its excitation energy is obtained to be small (36 keV), relatively close to experiment (1.7 keV).

(iii) We note that the magnetic moment obtained for the $3/2^{\langle +\rangle}$ ground state in the reflection-unconstrained calculation takes the value of $-0.248$ $\mu_N$ which considerably underestimates in absolute value the experimental value of $-0.404(2)$ $\mu_N$ \cite{Stone_2005}. On the other hand the value of $-0.326$ $\mu_N$ obtained in the symmetric solution for the $3/2^{+}$ state is closer to the experiment.

(iv) The above overall results show that by moving from $^{229}$Th to $^{227}$Ra under the condition of reflection asymmetry, the $5/2^{+}$ and $3/2^{+}$ orbitals exchange their roles in the ground and first excited q.p. state but still remain close to one another. The mutual rearrangement of these two orbitals is similar to what was observed in $^{227}$Th and $^{231}$Th.

We now turn to the case of  $^{231}$U, where both the reflection-symmetric and reflection-unconstrained calculations give the low-lying q.p. excitations with $5/2^{-}$, $5/2^{+}$, $7/2^{-}$ and $3/2^{+}$, as seen in Table~\ref{tab:neighbors} and Fig.~\ref{fig_Ra227_U231}(c)-(d). Here we can make the following comments:

(i) We see that in both reflection-symmetric and octupole calculations, the ground state is obtained with $K^{\pi}=5/2^{+}$ and not with the $K^{\pi}=5/2^{-}$ quantum number adopted in the experimental database \cite{ensdf}. However, as noted above, the very recent measurement \cite{Roux_EPJA2024} suggests that the ground state is based on the neutron 5/2$^{+}$[633] orbital thus corroborating our result. Our first excited state is $K^{\pi}=3/2^{+}$, then comes $K^{\pi}=5/2^{-}$ and finally $K^{\pi}=7/2^{-}$. We note that in both cases the obtained $K^{\pi}=5/2^{+}$ GS and the first excited $K^{\pi}=3/2^{+}$ state remain rather close to one another (see Fig.~\ref{fig_Ra227_U231}).

(ii) Except for $K^{\pi}=5/2^{\langle -\rangle}$ the calculations provide lower total energy in the case of reflection-unconstrained solutions (see Table~\ref{tab:neighbors}). Thus, the octupole ground state $K^{\pi}=5/2^{\langle +\rangle}$ and the $K^{\pi}=3/2^{\langle +\rangle}$ excitation appear slightly favored by 43 keV and 29 keV, respectively, with respect to the reflection symmetric case. At the same time, the $K^{\pi}=7/2^{\langle -\rangle}$ state goes deeper by 769 keV with a large $\beta_{3}=0.151$. For $K^{\pi}=5/2^{\langle -\rangle }$ the reflection-unconstrained calculation does not differ from the symmetric solution with $\beta_{3}=0$.

(iii) We notice that in most of the considered states, both calculations (reflection-symmetric and octupole) give quite large values for the collective gyromagnetic factor $g_R$ exceeding the phenomenological $Z/A=0.398$ ratio for $^{231}$U, while usually they would be expected to be smaller. Further detailed work is needed to understand this result.

Finally, we would like to note that we refrain from comparing in our analysis the theoretical excited state energies with experimental values for bandheads with the same angular momentum, which are for some cases available and might show considerable energy differences. The main reason is that in all cases the excited bandheads, especially those at relatively high energy of few hundreds of keV, are coupled to other dynamical modes---collective vibrations and rotations with possible Coriolis mixing effects, as well---which are not included in the present theoretical modeling. Presently the only possible way to better describe available excited states could be through variation of the pairing parameters trying to adjust them to reproduce the lowest one-q.p. excitation, in particular the $^{229m}$Th isomer, instead of relying on the adjustment to the core moment of inertia. However, such an attempt is beyond the scope of this work.

Summarizing the result of calculations in the $^{227,231}$Th isotopes and $^{227}$Ra, $^{231}$U isotones, we remark that the pair of $K^{\pi}=5/2^{+}$ and $K^{\pi}=3/2^{+}$ orbitals sustainedly plays a role in the formation of low-lying excitations in the nuclei around $^{229}$Th. This is illustrated in Fig.~\ref{fig_allnuc} where we collect and compare the low-lying one-neutron q.p. levels obtained for all considered nuclei in the case of reflection-unconstrained HFBCS calculations. We also note that in our calculations  the two orbitals mentioned above, together with the few other neighboring orbitals considered, keep the same Nilsson quantum numbers in the leading components of the corresponding s.p. wave functions as in $^{229}$Th (see Tables \ref{tab:th229reflsym} and \ref{tab:th229oct}). The observed evolution of the intrinsic structure in the vicinity of $^{229}$Th suggests that the same microscopic mechanism for formation of low-energy excitations, involving s.p. orbital-crossing or quasi-degeneracy, may govern in combination with other degrees of freedom, such as collective vibrations and rotations, the appearance of similar low-energy isomeric states in the same mass region. Furthermore, a search for the appearance of similar intrinsic-structure conditions in other mass regions seems to be worth doing.

\section{Conclusion and perspectives}
\label{over}

We applied a Skyrme HFBCS approach to study the microscopic grounds of the $^{229m}$Th isomer-formation mechanism and its possible manifestation in other nuclei. By properly handling the pairing, moment-of-inertia and quadrupole-octupole shape characteristics inherent to actinide nuclei we expanded stepwise the study over the even-even core and core-plus-particle intrinsic structures of $^{229}$Th and its closest odd-isotope $^{227}$Th, $^{231}$Th and odd-isotone $^{227}$Ra, $^{231}$U neighbors. Focusing on the isomer-forming s.p. orbitals and their neighboring orbitals, we examined the related low-lying q.p. excitations and the deformation and intrinsic moment characteristics in $^{229}$Th and around. We reach the following conclusions:

(i) The model calculation with the Skyrme SIII parametrization provides the necessary orbitals in the s.p. spectrum of the even-even core nucleus $^{228}$Th and subsequently the correct order of the ground $K^{\pi}=5/2^{+}$ and first excited $K^{\pi}=3/2^{+}$ states in $^{229}$Th promoting the latter as the model candidate for the isomer in both cases of reflection symmetry and octupole deformation.

(ii) The reflection-unconstrained (octupole) solutions provide lower total energy in all considered states of $^{229}$Th compared to the reflection-symmetric solutions. On the other hand, the latter provide lower $3/2^{+}$ excitation energy for $^{229}$Th, however for all considered cases  orders of magnitude larger than the 8 eV experimental isomer energy. Allegedly, eV accuracy is out of reach of current nuclear models.
Within the typically nuclear energy scale accuracy, we adopt the $3/2^{+}$ energy of about 40 keV from the octupole solution in Table~\ref{tab:th229oct} as the current SIII-HFBCS prediction for the $^{229m}$Th isomer.

(iii) The reflection-symmetric solution significantly overestimates the experimental $^{229}$Th GS magnetic moment while reproducing the IS magnetic moment within the experimental uncertainties. On the other hand, the octupole solution approaches the GS magnetic moment from above while slightly overestimating the IS magnetic moment. Overall, this result is very similar to the corresponding magnetic moment descriptions obtained in the CQOM-DSM approach \cite{Minkov_Palffy_PRL_2017,Minkov_Palffy_PRC_2021} and in earlier DSM calculations \cite{Chasman1977}. This consistency suggests that the  Skyrme-SIII HFBCS calculation may provide the relevant s.p. wave functions capable of predicting reasonably well the  $B(M1)$ radiative decay rate of $^{229m}$Th.

(iv) The analysis of the spacings between the lowest excited bandheads obtained in $^{229}$Th shows a stability of the model predictions for the quadrupole and octupole moments, deformation parameters and magnetic moments as well, compared to energy-level shifts of tens of keVs caused by the variation in the pairing strengths. It follows that within such limits of $3/2^{+}$ energy-level variation the obtained HFBCS predictions for the considered observables may provide reasonable qualitative characterization of the $^{229m}$Th isomer-formation conditions. Moreover, this suggests that further model calculations involving the B(M1) transition rates could also carry valuable microscopic information about the $3/2^{+}$ isomer-decay regardless of the keV-limited accuracy.

(v) The application of the SIII-HFBCS approach to the odd isotope and isotone neighbors of $^{229}$Th suggests a plausible evolution of the considered excitations in the actinide region of the nuclear chart. In $^{227}$Ra and $^{231}$Th the calculations suggest $3/2^{+}$ as the ground state, while setting $5/2^{+}$ as the first excited state, thus exchanging their roles as compared to $^{229}$Th. At the same time  $^{231}$U presents the same level order as  $^{229}$Th, with $5/2^{+}$ being the GS and $3/2^{+}$ appearing as the first excitation. In $^{227}$Th the calculations indicate a strong octupole deformation which essentially rearranges the levels pushing both $5/2^{\langle +\rangle}$ and $3/2^{\langle +\rangle}$ upper in energy and setting $1/2^{\langle +\rangle}$ as the GS in agreement with experiment. An interesting byproduct of our calculations is that $^{229}$Th appears at the border of octupole deformation in the series of odd-mass Thorium isotopes. The observed ``migration'' of other neighboring states, such as $5/2^{-}$ and $7/2^{-}$, around suggests a potential rearrangement of the s.p. orbitals, including crossing and quasi-degeneracy, in a way providing conditions for the appearance of yet not experimentally observed extremely low-energy excitations.

The main conclusion of the study is that the present Skyrme EDF approach gives us a basis for a deeper microscopic understanding of the mechanism which governs the formation of low and eventually extremely low-lying nuclear excitations. Our results point on the good spectroscopic quality of the SIII interaction in $^{229}$Th and its odd-mass neighbors. In this aspect it would be also interesting to test other Skyrme parametrizations, especially such with effective mass close to one, known to be successful in reproducing single-particle spectra around the Fermi level. At the same time the further practical application of the approach on a quantitative level (apart from the possible tuning of pairing constants) requires a few developments in different directions. First, the inclusion of the Coriolis coupling between the odd nucleon and the core (implying a sophisticated evaluation of the core moment of inertia and relevant intrinsic matrix element) with the $K$-mixing effect allowing for the $M1$ decay and additional coupling to collective rotation and/or vibration modes allowing for electric $E2$ and $E1$ or $E3$ transitions. Further, detailed implementation of nuclear shape effects, parity projection, moments of inertia and gyromagnetic characteristics within a unified particle-core framework will be necessary to reach a quantitative description and prediction of energy and electromagnetic properties of $^{229m}$Th on a level of accuracy comparable with the more phenomenological schemes such as the CQOM-DSM. This development could be the subject of further work.

\section*{ACKNOWLEDGMENTS} The authors would like to thank P.~G.~Reinhard for fruitful discussions. This work is supported by the Bulgarian National Science Fund (BNSF) under Contract No. KP-06-N48/1. N.M. acknowledges with thanks the use of the MPIK-Heidelberg computer cluster allowed through a collaboration with the Division of Prof. Dr. Klaus Blaum.
AP gratefully acknowledges support from the Deutsche Forschungsgemeinschaft (DFG, German Science Foundation) in the framework of the Heisenberg Program (PA 2508/3-1) and from the ThoriumNuclearClock project that has received funding from the European Research Council (ERC) under the European Union’s Horizon
2020 research and innovation programme (Grant Agreement
No. 856415).

\appendix

\section{Minimal scheme of time-odd terms in Skyrme energy-density
  functional}

First, we recall here the expressions of the central, density-dependent and
spin-orbit parts of the energy-density functional, omitting the coordinate
$\vect r$ dependence of local densities to shorten the expressions
\begin{subequations}
\begin{align}
  \label{H_c}
  \mathcal H_{\rm c}(\mathbf r) = &
B_1 \, \rho^2 + B_{10} \, \vect s^2 + B_3 \, (\rho \,\tau - \vect j^2)
  \nonumber \\
& + B_{14} \, (\overleftrightarrow{\mbox{J}}^2 - \vect s \cdot \vect T)
+ B_5 \,
\rho\,\Delta\rho + B_{18} \, \vect s \cdot \Delta\vect s \nonumber \\
& + \sum_{q=n,p} \Big\{ B_2 \,\rho_{q}^2 + B_{11}\,\vect s_{q}^2 +
  B_4 \, (\rho_{q}\tau_{q} - \vect j_{q}^2) \nonumber \\
& \phantom{+ \sum_{q} \Big\{ }
+ B_{15}
  (\overleftrightarrow{\mbox{J}_{q}}^2 - \vect s_{q} \cdot \vect T_{q}) + B_6 \,
  \rho_{q}\,\Delta\rho_{q} \nonumber \\
& \phantom{+ \sum_{q} \Big\{ } + B_{19} \, \vect s_{q} \cdot \Delta\vect s_{q}\Big\} \:, \\
\label{H_DD}
\mathcal H_{\rm DD}(\vect r) = &
\rho^{\alpha}\,\Big[B_7 \,\rho^2 + B_{12} \, \vect s^2 + \sum_{q=n,p} \big(B_8 \,
\rho_{q}^2 + B_{13} \, \vect s_{q}^2 \big)\Big] \:,\\[-0.25cm]
\label{H_so}
\mathcal H_{\rm s.o.}(\vect r) = & B_9\,\Big[
\rho \, \bm{\nabla}\cdot\vect J +
\vect j \cdot \bm{\nabla} \times\vect s \nonumber \\
& \phantom{B_9\,\Big[} + \sum_{q=n,p} \Big(\rho_{q} \,
\bm{\nabla}\cdot\vect J_{q} + \vect j_{q} \cdot \bm{\nabla}\times\vect s_{q}\Big)
\Big] \:,
\end{align}
\end{subequations}
where the index $q$ in local densities is the charge state ($q=n$ for neutrons or $q=p$ for protons). In Eq.~(\ref{H_c}), $\overleftrightarrow{\mbox{J}_{q}}^2 = J_{q,\mu\nu}J_{q}^{\mu\nu}$ (using Einstein's summation convention) and in Eq.~(\ref{H_so}) $\vect J_{q}$ denotes the antisymmetric part of the spin-current tensor $J^{\mu\nu}_{q}$~\cite{Engel75}. The coupling constants $B_i$ are related to the Skyrme parameters by expressions given, e.g., in Ref.~\cite{Hellemans12}. The time-even local densities $\rho(\vect r)$, $\tau(\vect r)$ and $J_{\mu\nu}(\vect r)$ appearing in Eqs.~(\ref{H_c}) to (\ref{H_so}) are defined in terms of the spinor wave functions $[\psi_k](\vect r)$ of the Hartree--Fock basis states by
\begin{subequations}
  \begin{align}
    \label{rho}
\rho(\vect r) & = \sum_k v_k^2 \, [\psi_k]^{\dagger}(\vect r)[\psi_k](\vect r) \:, \\
\tau(\vect r) & = \sum_k v_k^2 \, \Big(\bm{\nabla}[\psi_k]^{\dagger}(\vect r)\Big) \cdot \bm{\nabla}[\psi_k](\vect r) \:, \\
J_{\mu\nu}(\vect r) & = \frac{1}{2i}\sum_k v_k^2 \, \Big\{
[\psi_k]^{\dagger}(\vect r) \, \sigma_{\nu} \partial_{\mu}[\psi_k](\vect r)
\nonumber \\
& \phantom{=\frac{1}{2i}\sum_k v_k^2\Big\{}
-\Big(\partial_{\mu}[\psi_k]^{\dagger}(\vect r)\Big) \,
\sigma_{\nu}[\psi_k](\vect r) \Big\} \: ,
\end{align}
\end{subequations}
whereas the relevant time-odd densities $\vect s(\vect r)$, $T_{\mu}(\vect r)$ and $\vect j(\vect r)$ respectively read
\begin{subequations}\begin{align}
\vect s(\vect r) & = \sum_k v_k^2 \, [\psi_k]^{\dagger}(\vect r) \,
\bm{\sigma}[\psi_k](\vect r) \:, \\
T_{\mu}(\vect r) & = \sum_k v_k^2 \,
\Big(\bm{\nabla}[\psi_k]^{\dagger}(\vect r)\Big)
\cdot \sigma_{\mu} \bm{\nabla}[\psi_k](\vect r) \:, \\
\vect j(\vect r) & = \frac{1}{2i}\sum_k v_k^2 \, \Big\{
\Big(\bm{\nabla}[\psi_k]^{\dagger}(\vect r)\Big)[\psi_k](\vect r)
\nonumber \\
& \phantom{=\frac{1}{2i}\sum_k v_k^2\Big\{}
- [\psi_k]^{\dagger}(\vect r)\bm{\nabla}[\psi_k](\vect r)
\Big\} \:.
\end{align}\end{subequations}
It is worth noting that the summation runs over all the Hartree--Fock basis without assuming any specific relation between the wave functions of canonically conjugate states because of time-reversal symmetry breaking at the one-body level.

The Hartree--Fock one-body Hamiltonian $\hat h_{\rm HF}^{(q)}$ takes the following form in coordinate representation for the charge state $q$
\begin{align}
\elmx{\mathbf r}{\hat h_{\rm HF}^{(q)}}{\psi_k} = &
-\bm{\nabla} \cdot \Big[\Big(\frac{\hbar^2}{2m^*_{q}(\mathbf r)} +
\vect C_{q}(\vect r) \cdot \bm{\sigma} \Big)\bm{\nabla}
[\psi_k](\mathbf r)\Big] \nonumber \\
& + \Big(U_{q}(\vect r) +\delta_{q\, p}\,V_{\rm
  Coul}(\vect r)\Big) [\psi_k](\mathbf r) \nonumber \\
& +i \, \mathbf{W}_{q}(\mathbf r) \cdot \Big(\bm{\sigma} \times
\bm{\nabla} [\psi_k](\mathbf r) \Big) \nonumber \\
& - i \Big[ W_{q}^{\mu \nu}(\vect r) \sigma_{\nu} \partial_{\mu}
[\psi_k](\mathbf r) \nonumber \\
& + \nabla_{\mu} \Big( W_{q}^{\mu  \nu}(\vect r)
\sigma_{\nu} [\psi_k](\mathbf r) \Big) \Big] \nonumber \\
& - i \, \vect A_{q}(\vect r) \cdot \bm{\nabla}[\psi_k](\mathbf
r) + \vect S_{q}(\vect r) \cdot \bm{\sigma}[\psi_k](\mathbf r)  \:.
\label{h_HF}
\end{align}
The time-even fields $m^*_{q}(\vect r)$, $U_{q}(\vect r)$, $V_{\rm Coul}(\vect r)$, $\mathbf{W}_{q}(\vect r)$ and $W_{q}^{\mu\nu}(\vect r)$ denote the effective-mass field, the central-plus-density-dependent field, the Coulomb field, the spin-orbit field and the spin-current field respectively, whereas $\vect S_{q}(\vect r)$, $\vect A_{q}(\vect r)$ and $\vect C_{q}(\vect r)$ are time-odd fields. The expressions of all the fields can be found in, e.g., Refs.~\cite{Hellemans12,PRC15_hfbcs}, but we recall here those relevant for the ``minimal scheme'' definition
\begin{subequations}\begin{align}
\label{Wqmunu}
W_{q}^{\mu \nu}(\vect r) & = B_{14} J^{\mu\nu} + B_{15}J_{q}^{\mu\nu} \:, \\
\label{Sq}
\vect S_{q}(\vect r) & = 2\big(B_{10}+B_{12} \, \rho^{\alpha}) \vect s +
2\big(B_{11} + B_{13} \, \rho^{\alpha}\big) \vect s_{q} \nonumber \\
& \phantom{=} - B_9 \, \bm{\nabla} \times (\vect j + \vect j_{q})
- B_{14} \, \vect T - B_{15} \, \vect T_{q}  \nonumber \\
& \phantom{=} + 2 \, \Big( B_{18} \, \bm{\Delta} \vect s + B_{19}
\, \bm{\Delta} \vect s_{q}\Big) \:, \\
\label{Aq}
\vect A_{q}(\vect r) & = -2\big(B_{3} \, \vect j + B_{4} \, \vect j_{q}\big)
+ B_9 \, \bm{\nabla} \times (\vect s + \vect s_{q}) \:, \\
\label{Cq}
\vect C_{q}(\vect r) & = -B_{14} \, \vect s - B_{15} \, \vect s_{q} \:.
\end{align}\end{subequations}

In this context, the so-called ``minimal scheme'' for the Skyrme SIII parametrization is then defined by the neglect, in the energy density functional as well as in the Hartree--Fock potential, of
\begin{itemize}
\item the $B_{14}$ and $B_{15}$ terms because they were not included
  in the fit of Skyrme SIII parameters;
\item the $B_{18}$ and $B_{19}$ terms because they may yield
  finite-size spin instabilities according to Ref.~\cite{Hellemans12}.
\end{itemize}
Therefore, the local densities $W_q^{\mu\nu}(\vect r)$ and $\vect T_q(\vect r)$, as well as the Laplacian of the spin density, $\bm{\Delta}\vect s_q(\vect r)$, are not involved in our calculations with SIII. As a consequence, the field $\vect C_q(\vect r)$, which adds a spin-dependent contribution to the effective-mass field, and the field $W_q^{\mu\nu}(\vect r)$, introducing a tensor contribution to the spin-orbit potential, disappear from the Hartree--Fock Hamiltonian in Eq.~(\ref{h_HF}). The only time-odd contributions to $\hat h_{\rm HF}^{(q)}$ thus come from the current field $\vect A_q(\vect r)$ and the spin field $\vect S_q(\vect r)$ reduced to
\begin{align}
  \vect S_{q}(\vect r) \approx & \, 2\big(B_{10}+B_{12} \, \rho^{\alpha})
  \vect s +
  2\big(B_{11} + B_{13} \, \rho^{\alpha}\big) \vect s_{q} \nonumber \\
  & - B_9 \, \bm{\nabla} \times (\vect j + \vect j_{q}) \,.
\end{align}

\section{Quadrupole and octupole moments and deformation
  parameters in the HFBCS calculation}

In the HFBCS algorithm of Ref.~\cite{PRC15_hfbcs} the quadrupole
deformation parameter $\beta$ can be determined in two ways, through
the mass $(m)$ or charge ($p$ - proton) quadrupole moments and radii, as
follows:
\begin{eqnarray}
\beta_{2}^{m} & = &\sqrt{\frac{\pi}{5}}\frac{\langle
  \widehat{Q}_{20}^{m}\rangle}{A\langle r^{2}_{m}\rangle}\ ,
\label{beta2Q2m}\\
\beta_{2}^{p} & = &\sqrt{\frac{\pi}{5}}\frac{\langle
  \widehat{Q}^{p}_{20}\rangle}{Z\langle r^{2}_{p}\rangle} \ ,
\label{beta2Q2p}
\end{eqnarray}
where $\widehat{Q}_{20}^{m}$ ($\widehat{Q}^{p}_{20}$) and  $\langle
r^{2}_{m}\rangle$ ($\langle r^{2}_{p}\rangle$) are the mass (charge)
quadrupole momenta and mean quadratic radii calculated in the code as
\begin{eqnarray}
\langle \widehat{Q}_{20}^{m}\rangle & =& \int d\mathbf{r}\:\rho
(\mathbf{r})\: (3z^{2}-\mathbf{r}^{2})\ ,\label{Q20m}\\
\langle \widehat{Q}_{20}^{p}\rangle & =& \int d\mathbf{r}\:\rho_p
(\mathbf{r})\: (3z^{2}-\mathbf{r}^{2})\ ,
\label{Q20p}
\end{eqnarray}
and
\begin{eqnarray}
\langle r^{2}_{m}\rangle &=&\frac{1}{A}\int d\mathbf{r}\:\rho (\mathbf{r})\:
\mathbf{r}^{2} \ ,\\
\langle r^{2}_{p}\rangle &=&\frac{1}{Z}\int d\mathbf{r}\:\rho_p (\mathbf{r})\:
\mathbf{r}^{2} \ ,
\end{eqnarray}
where $\rho(\mathbf{r})$ and $\rho_p(\mathbf{r})$ are the total (mass) and proton (charge) densities, respectively, and $\mathbf{r}^{2}=z^2+x^2+y^2$. The mass and charge densities $\rho(\mathbf{r})$ and $\rho_p(\mathbf{r})$ are basic ingredients of the HFBCS solution (see Eq.~(\ref{rho}) for the definition of the former, whereas the restriction in the summation to proton states gives $\rho_p(\mathbf{r})$). In the calculation in formula (\ref{beta2Q2p}) the charge mean quadratic radii $\langle r^{2}_{p}\rangle$ are corrected by the addition of the value $a^{2}=(0.85)^2$ fm$^2$ taking into account the proton radius.

The octupole deformation is calculated from the mass and charge
octupole moments as
\begin{eqnarray}
\beta_{3}^{m}&=&\frac{4\pi}{3AR_{0}^{3}} \langle
\widehat{\overline{Q}}_{30}^{m}\rangle\ , \label{beta3Q30mbar}\\
\beta_{3}^{p}&=&\frac{4\pi}{3ZR_{0}^{3}} \langle
\widehat{\overline{Q}}_{30}^{p}\rangle\ , \label{beta3Q30pbar}
\end{eqnarray}
where
\begin{eqnarray}
\langle \widehat{\overline{Q}}_{30}^{m}\rangle & =&
\frac{1}{2}\sqrt{\frac{7}{4\pi}}\int d\mathbf{r}\:\rho (\mathbf{r})\:
z(2z^{2}-3x^2-3y^2),\\
\langle \widehat{\overline{Q}}_{30}^{p}\rangle & =&
\frac{1}{2}\sqrt{\frac{7}{4\pi}}\int d\mathbf{r}\:\rho_p (\mathbf{r})\:
z(2z^{2}-3x^2-3y^2).
\end{eqnarray}
Here, $R_{0}=r_{0}A^{1/3}$ with $r_{0}=1.2$ fm.

The above expressions (\ref{beta2Q2m}), (\ref{beta2Q2p}) and
(\ref{beta3Q30mbar}), (\ref{beta3Q30pbar}) illustrate the particular
way in which the quadrupole and octupole moments and deformation
parameters are calculated in the present HFBCS algorithm.  They can
be also written in a more unified form and also related to different forms
used in other works.

Thus, we remark that Eqs. (\ref{beta2Q2m}) and (\ref{beta2Q2p}) can be
approximately written in terms of $R_{0}$ as in the case of
Eqs.~(\ref{beta3Q30mbar}) and (\ref{beta3Q30pbar}). Considering that
for small quadrupole deformations one can approximate the mean square
radius $\langle r^{2}\rangle$, similarly to the radius of a system of
$A$ particles in a spherical harmonic oscillator potential, as (see
Eq.~(2.11) on p. 41 of Ref.~\cite{RS80})
\begin{eqnarray}
\langle r^{2}\rangle = \frac{1}{A}\sum_{i=1}^{A}\langle
r^{2}\rangle_{i}\simeq \frac{3}{5}\left(r_{0}A^{1/3}\right)^{2} =
\frac{3}{5}R_{0}^{2}\ ,
\end{eqnarray}
and substituting it into Eqs. (\ref{beta2Q2m}) and (\ref{beta2Q2p})
one gets
\begin{eqnarray}
\beta_{2}^{m} &=&\frac{\sqrt{5\pi}}{3AR_{0}^{2}} \langle
\widehat{Q}_{20}^{m}\rangle = \frac{\sqrt{5\pi}}{3r_{0}^{2}A^{5/3}}
\langle \widehat{Q}_{20}^{m}\rangle \label{betaQ20mR0}\ ,\\
\beta_{2}^{p}
&=&\frac{\sqrt{5\pi}}{3ZR_{0}^{2}}\langle\widehat{Q}_{20}^{p}\rangle =
\frac{\sqrt{5\pi}}{3Zr_{0}^{2}A^{2/3}}\langle\widehat{Q}_{20}^{p}\rangle\ .
\label{betaQ20pR0}
\end{eqnarray}
Eq.~(\ref{betaQ20mR0}) corresponds to Eq.~(34) in \cite{Chen21} and
Eq.~(3) of \cite{Nom21}, while Eq.~(\ref{betaQ20pR0}) corresponds to
Eq.~(1.72) in Ref.~\cite{RS80} and, e.g., Eq.~(11) in
Ref.~\cite{Dutta91} up to the first order of $\beta_{2}$.

Regarding the octupole moments and deformation parameters, if we
consider
\begin{eqnarray}
\langle \widehat{\overline{Q}}_{30}\rangle =
\frac{1}{2}\sqrt{\frac{7}{4\pi}}\langle \widehat{Q}_{30}\rangle,
\end{eqnarray}
with
\begin{eqnarray}
\langle \widehat{Q}_{30}\rangle  = \int d\mathbf{r}\:\rho
(\mathbf{r})\: z(2z^{2}-3x^2-3y^2),
\label{Q30}
\end{eqnarray}
one obtains
\begin{eqnarray}
\beta_{3}^{m}&=&\frac{\sqrt{7\pi}}{3AR_{0}^{3}} \langle
\widehat{Q}_{30}^{m}\rangle \label{beta3Q30m}\ ,\\
\beta_{3}^{p}&=&\frac{\sqrt{7\pi}}{3ZR_{0}^{3}} \langle
\widehat{Q}_{30}^{p}\rangle\ .
\label{beta3Q30p}
\end{eqnarray}
Thus, Eq.~\ref{beta3Q30mbar} corresponds to Eq.~(34) in \cite{Chen21}
and Eq.~(4) of \cite{Nom21}, while Eq.~(\ref{beta3Q30pbar})
corresponds to Eq.~(6) in Ref.~\cite{BN96}.

Also, we remark that the expectation values of the quadrupole and
octupole operators in Eqs.~(\ref{Q20m}), (\ref{Q20p}) and (\ref{Q30}),
respectively, correspond to Eqs.~(1) and (2) in \cite{Nom21},
respectively, as well as to  those in Eq.~(35) of Ref.~\cite{Chen21}.

Finally, unifying Eq.~(\ref{betaQ20mR0}) with Eq.~(\ref{beta3Q30m}) and
Eq.~(\ref{betaQ20pR0}) with Eq.~(\ref{beta3Q30p}) we obtain
\begin{eqnarray}
\beta_{\lambda}^{m}&=&\frac{\sqrt{(2\lambda
    +1)\pi}}{3AR_{0}^{\lambda}} \langle \widehat{Q}_{\lambda
  0}^{m}\rangle\ , \label{betalamm}\\
\beta_{\lambda}^{p}&=&\frac{\sqrt{(2\lambda
    +1)\pi}}{3ZR_{0}^{\lambda}} \langle \widehat{Q}_{\lambda
  0}^{p}\rangle\ ,
\label{betalamp}
\end{eqnarray}
with $\lambda =2,3$, where Eq.~(\ref{betalamp}) corresponds to
Eq.~(2.23) in Ref.~\cite{LC88}.

  \bibliography{refs}

\begin{thebibliography}{82}
\expandafter\ifx\csname natexlab\endcsname\relax\def\natexlab#1{#1}\fi
\expandafter\ifx\csname bibnamefont\endcsname\relax
  \def\bibnamefont#1{#1}\fi
\expandafter\ifx\csname bibfnamefont\endcsname\relax
  \def\bibfnamefont#1{#1}\fi
\expandafter\ifx\csname citenamefont\endcsname\relax
  \def\citenamefont#1{#1}\fi
\expandafter\ifx\csname url\endcsname\relax
  \def\url#1{\texttt{#1}}\fi
\expandafter\ifx\csname urlprefix\endcsname\relax\def\urlprefix{URL }\fi
\providecommand{\bibinfo}[2]{#2}
\providecommand{\eprint}[2][]{\url{#2}}

\bibitem[{\citenamefont{{E. Peik} and {Chr. Tamm}}(2003)}]{Peik_Clock_2003}
\bibinfo{author}{\bibnamefont{{E. Peik}}} \bibnamefont{and}
  \bibinfo{author}{\bibnamefont{{Chr. Tamm}}}, \bibinfo{journal}{Europhys.
  Lett.} \textbf{\bibinfo{volume}{61}}, \bibinfo{pages}{181}
  (\bibinfo{year}{2003}),
  \urlprefix\url{https://doi.org/10.1209/epl/i2003-00210-x}.

\bibitem[{\citenamefont{Campbell et~al.}(2012)\citenamefont{Campbell, Radnaev,
  Kuzmich, Dzuba, Flambaum, and Derevianko}}]{Campbell_Clock_2012}
\bibinfo{author}{\bibfnamefont{C.~J.} \bibnamefont{Campbell}},
  \bibinfo{author}{\bibfnamefont{A.~G.} \bibnamefont{Radnaev}},
  \bibinfo{author}{\bibfnamefont{A.}~\bibnamefont{Kuzmich}},
  \bibinfo{author}{\bibfnamefont{V.~A.} \bibnamefont{Dzuba}},
  \bibinfo{author}{\bibfnamefont{V.~V.} \bibnamefont{Flambaum}},
  \bibnamefont{and}
  \bibinfo{author}{\bibfnamefont{A.}~\bibnamefont{Derevianko}},
  \bibinfo{journal}{Phys. Rev. Lett.} \textbf{\bibinfo{volume}{108}},
  \bibinfo{pages}{120802} (\bibinfo{year}{2012}),
  \urlprefix\url{https://link.aps.org/doi/10.1103/PhysRevLett.108.120802}.

\bibitem[{\citenamefont{Peik and Okhapkin}(2015)}]{Peik_Clock_2015}
\bibinfo{author}{\bibfnamefont{E.}~\bibnamefont{Peik}} \bibnamefont{and}
  \bibinfo{author}{\bibfnamefont{M.}~\bibnamefont{Okhapkin}},
  \bibinfo{journal}{Comptes Rendus Physique} \textbf{\bibinfo{volume}{16}},
  \bibinfo{pages}{516 } (\bibinfo{year}{2015}),
  \urlprefix\url{http://www.sciencedirect.com/science/article/pii/S1631070515000213}.

\bibitem[{\citenamefont{Flambaum}(2006)}]{flambaum_2006}
\bibinfo{author}{\bibfnamefont{V.~V.} \bibnamefont{Flambaum}},
  \bibinfo{journal}{Phys. Rev. Lett.} \textbf{\bibinfo{volume}{97}},
  \bibinfo{pages}{092502} (\bibinfo{year}{2006}),
  \urlprefix\url{https://link.aps.org/doi/10.1103/PhysRevLett.97.092502}.

\bibitem[{\citenamefont{Fadeev et~al.}(2020)\citenamefont{Fadeev, Berengut, and
  Flambaum}}]{Fadeev_2020}
\bibinfo{author}{\bibfnamefont{P.}~\bibnamefont{Fadeev}},
  \bibinfo{author}{\bibfnamefont{J.~C.} \bibnamefont{Berengut}},
  \bibnamefont{and} \bibinfo{author}{\bibfnamefont{V.~V.}
  \bibnamefont{Flambaum}}, \bibinfo{journal}{Phys. Rev. A}
  \textbf{\bibinfo{volume}{102}}, \bibinfo{pages}{052833}
  (\bibinfo{year}{2020}),
  \urlprefix\url{https://link.aps.org/doi/10.1103/PhysRevA.102.052833}.

\bibitem[{\citenamefont{Peik et~al.}(2021)\citenamefont{Peik, Schumm,
  Safronova, P\'alffy, Weitenberg, and Thirolf}}]{peik2021nuclear}
\bibinfo{author}{\bibfnamefont{E.}~\bibnamefont{Peik}},
  \bibinfo{author}{\bibfnamefont{T.}~\bibnamefont{Schumm}},
  \bibinfo{author}{\bibfnamefont{M.}~\bibnamefont{Safronova}},
  \bibinfo{author}{\bibfnamefont{A.}~\bibnamefont{P\'alffy}},
  \bibinfo{author}{\bibfnamefont{J.}~\bibnamefont{Weitenberg}},
  \bibnamefont{and} \bibinfo{author}{\bibfnamefont{P.~G.}
  \bibnamefont{Thirolf}}, \bibinfo{journal}{Quantum Science and Technology}
  \textbf{\bibinfo{volume}{6}}, \bibinfo{pages}{034002} (\bibinfo{year}{2021}).

\bibitem[{\citenamefont{Tsai et~al.}(2023)\citenamefont{Tsai, Eby, and
  Safronova}}]{tsai2023direct}
\bibinfo{author}{\bibfnamefont{Y.-D.} \bibnamefont{Tsai}},
  \bibinfo{author}{\bibfnamefont{J.}~\bibnamefont{Eby}}, \bibnamefont{and}
  \bibinfo{author}{\bibfnamefont{M.~S.} \bibnamefont{Safronova}},
  \bibinfo{journal}{Nature Astronomy} \textbf{\bibinfo{volume}{7}},
  \bibinfo{pages}{113} (\bibinfo{year}{2023}).

\bibitem[{\citenamefont{Beck et~al.}(2007)\citenamefont{Beck, Becker,
  Beiersdorfer, Brown, Moody, Wilhelmy, Porter, Kilbourne, and
  Kelley}}]{Beck_78eV_2007}
\bibinfo{author}{\bibfnamefont{B.~R.} \bibnamefont{Beck}},
  \bibinfo{author}{\bibfnamefont{J.~A.} \bibnamefont{Becker}},
  \bibinfo{author}{\bibfnamefont{P.}~\bibnamefont{Beiersdorfer}},
  \bibinfo{author}{\bibfnamefont{G.~V.} \bibnamefont{Brown}},
  \bibinfo{author}{\bibfnamefont{K.~J.} \bibnamefont{Moody}},
  \bibinfo{author}{\bibfnamefont{J.~B.} \bibnamefont{Wilhelmy}},
  \bibinfo{author}{\bibfnamefont{F.~S.} \bibnamefont{Porter}},
  \bibinfo{author}{\bibfnamefont{C.~A.} \bibnamefont{Kilbourne}},
  \bibnamefont{and} \bibinfo{author}{\bibfnamefont{R.~L.}
  \bibnamefont{Kelley}}, \bibinfo{journal}{Phys. Rev. Lett.}
  \textbf{\bibinfo{volume}{98}}, \bibinfo{pages}{142501}
  (\bibinfo{year}{2007}),
  \urlprefix\url{https://link.aps.org/doi/10.1103/PhysRevLett.98.142501}.

\bibitem[{\citenamefont{Beck et~al.}(2009)\citenamefont{Beck, Wu, Beiersdorfer,
  Brown, Becker, Moody, Wilhelmy, Porter, Kilbourne, and
  Kelley}}]{Beck_78eV_2007_corrected}
\bibinfo{author}{\bibfnamefont{B.~R.} \bibnamefont{Beck}},
  \bibinfo{author}{\bibfnamefont{C.~Y.} \bibnamefont{Wu}},
  \bibinfo{author}{\bibfnamefont{P.}~\bibnamefont{Beiersdorfer}},
  \bibinfo{author}{\bibfnamefont{G.~V.} \bibnamefont{Brown}},
  \bibinfo{author}{\bibfnamefont{J.~A.} \bibnamefont{Becker}},
  \bibinfo{author}{\bibfnamefont{K.~J.} \bibnamefont{Moody}},
  \bibinfo{author}{\bibfnamefont{J.~B.} \bibnamefont{Wilhelmy}},
  \bibinfo{author}{\bibfnamefont{F.~S.} \bibnamefont{Porter}},
  \bibinfo{author}{\bibfnamefont{C.~A.} \bibnamefont{Kilbourne}},
  \bibnamefont{and} \bibinfo{author}{\bibfnamefont{R.~L.}
  \bibnamefont{Kelley}}, in \emph{\bibinfo{booktitle}{12th International
  Conference on Nuclear Reaction Mechanisms}} (\bibinfo{address}{Varenna,
  Italy}, \bibinfo{year}{2009}), vol. \bibinfo{volume}{LLNL-PROC-415170}.

\bibitem[{\citenamefont{Seiferle et~al.}(2019)}]{Seiferle_EnTh229m_2019}
\bibinfo{author}{\bibfnamefont{B.}~\bibnamefont{Seiferle}}
  \bibnamefont{et~al.}, \bibinfo{journal}{Nature (London)}
  \textbf{\bibinfo{volume}{573}}, \bibinfo{pages}{243} (\bibinfo{year}{2019}).

\bibitem[{\citenamefont{Yamaguchi et~al.}(2019)}]{Yamaguchi_EnTh229m_2019}
\bibinfo{author}{\bibfnamefont{A.}~\bibnamefont{Yamaguchi}}
  \bibnamefont{et~al.}, \bibinfo{journal}{Phys. Rev. Lett.}
  \textbf{\bibinfo{volume}{123}}, \bibinfo{pages}{222501}
  (\bibinfo{year}{2019}).

\bibitem[{\citenamefont{Sikorsky et~al.}(2020)\citenamefont{Sikorsky, Geist,
  Hengstler, Kempf, Gastaldo, Enss, Mokry, Runke, D{\"u}llmann, Wobrauschek
  et~al.}}]{Sikorsky2020}
\bibinfo{author}{\bibfnamefont{T.}~\bibnamefont{Sikorsky}},
  \bibinfo{author}{\bibfnamefont{J.}~\bibnamefont{Geist}},
  \bibinfo{author}{\bibfnamefont{D.}~\bibnamefont{Hengstler}},
  \bibinfo{author}{\bibfnamefont{S.}~\bibnamefont{Kempf}},
  \bibinfo{author}{\bibfnamefont{L.}~\bibnamefont{Gastaldo}},
  \bibinfo{author}{\bibfnamefont{C.}~\bibnamefont{Enss}},
  \bibinfo{author}{\bibfnamefont{C.}~\bibnamefont{Mokry}},
  \bibinfo{author}{\bibfnamefont{J.}~\bibnamefont{Runke}},
  \bibinfo{author}{\bibfnamefont{C.~E.} \bibnamefont{D{\"u}llmann}},
  \bibinfo{author}{\bibfnamefont{P.}~\bibnamefont{Wobrauschek}},
  \bibnamefont{et~al.}, \bibinfo{journal}{Phys. Rev. Lett.}
  \textbf{\bibinfo{volume}{125}}, \bibinfo{pages}{142503}
  (\bibinfo{year}{2020}),
  \urlprefix\url{https://journals.aps.org/prl/abstract/10.1103/PhysRevLett.125.142503}.

\bibitem[{\citenamefont{Kraemer et~al.}(2023)\citenamefont{Kraemer, Moens,
  Athanasakis-Kaklamanakis, Bara, Beeks, Chhetri, Chrysalidis, Claessens,
  Cocolios, Correia et~al.}}]{kraemer2023observation}
\bibinfo{author}{\bibfnamefont{S.}~\bibnamefont{Kraemer}},
  \bibinfo{author}{\bibfnamefont{J.}~\bibnamefont{Moens}},
  \bibinfo{author}{\bibfnamefont{M.}~\bibnamefont{Athanasakis-Kaklamanakis}},
  \bibinfo{author}{\bibfnamefont{S.}~\bibnamefont{Bara}},
  \bibinfo{author}{\bibfnamefont{K.}~\bibnamefont{Beeks}},
  \bibinfo{author}{\bibfnamefont{P.}~\bibnamefont{Chhetri}},
  \bibinfo{author}{\bibfnamefont{K.}~\bibnamefont{Chrysalidis}},
  \bibinfo{author}{\bibfnamefont{A.}~\bibnamefont{Claessens}},
  \bibinfo{author}{\bibfnamefont{T.~E.} \bibnamefont{Cocolios}},
  \bibinfo{author}{\bibfnamefont{J.~G.} \bibnamefont{Correia}},
  \bibnamefont{et~al.}, \bibinfo{journal}{Nature}
  \textbf{\bibinfo{volume}{617}}, \bibinfo{pages}{706} (\bibinfo{year}{2023}).

\bibitem[{\citenamefont{Tiedau et~al.}(2024)\citenamefont{Tiedau, Okhapkin,
  Zhang, Thielking, Zitzer, Peik, Schaden, Pronebner, Morawetz, De~Col
  et~al.}}]{PRL2024}
\bibinfo{author}{\bibfnamefont{J.}~\bibnamefont{Tiedau}},
  \bibinfo{author}{\bibfnamefont{M.~V.} \bibnamefont{Okhapkin}},
  \bibinfo{author}{\bibfnamefont{K.}~\bibnamefont{Zhang}},
  \bibinfo{author}{\bibfnamefont{J.}~\bibnamefont{Thielking}},
  \bibinfo{author}{\bibfnamefont{G.}~\bibnamefont{Zitzer}},
  \bibinfo{author}{\bibfnamefont{E.}~\bibnamefont{Peik}},
  \bibinfo{author}{\bibfnamefont{F.}~\bibnamefont{Schaden}},
  \bibinfo{author}{\bibfnamefont{T.}~\bibnamefont{Pronebner}},
  \bibinfo{author}{\bibfnamefont{I.}~\bibnamefont{Morawetz}},
  \bibinfo{author}{\bibfnamefont{L.~T.} \bibnamefont{De~Col}},
  \bibnamefont{et~al.}, \bibinfo{journal}{Phys. Rev. Lett.}
  \textbf{\bibinfo{volume}{132}}, \bibinfo{pages}{182501}
  (\bibinfo{year}{2024}),
  \urlprefix\url{https://link.aps.org/doi/10.1103/PhysRevLett.132.182501}.

\bibitem[{\citenamefont{von~der Wense et~al.}(2016)\citenamefont{von~der Wense,
  Seiferle, Laatiaoui, Neumayr, Maier, Wirth, Mokry, Runke, Eberhardt,
  D{\"u}llmann et~al.}}]{Wense_Nature_2016}
\bibinfo{author}{\bibfnamefont{L.}~\bibnamefont{von~der Wense}},
  \bibinfo{author}{\bibfnamefont{B.}~\bibnamefont{Seiferle}},
  \bibinfo{author}{\bibfnamefont{M.}~\bibnamefont{Laatiaoui}},
  \bibinfo{author}{\bibfnamefont{J.~B.} \bibnamefont{Neumayr}},
  \bibinfo{author}{\bibfnamefont{H.-J.} \bibnamefont{Maier}},
  \bibinfo{author}{\bibfnamefont{H.-F.} \bibnamefont{Wirth}},
  \bibinfo{author}{\bibfnamefont{C.}~\bibnamefont{Mokry}},
  \bibinfo{author}{\bibfnamefont{J.}~\bibnamefont{Runke}},
  \bibinfo{author}{\bibfnamefont{K.}~\bibnamefont{Eberhardt}},
  \bibinfo{author}{\bibfnamefont{C.~E.} \bibnamefont{D{\"u}llmann}},
  \bibnamefont{et~al.}, \bibinfo{journal}{Nature}
  \textbf{\bibinfo{volume}{533}}, \bibinfo{pages}{47} (\bibinfo{year}{2016}),
  ISSN \bibinfo{issn}{0028-0836}, \bibinfo{note}{article},
  \urlprefix\url{http://dx.doi.org/10.1038/nature17669}.

\bibitem[{\citenamefont{Seiferle et~al.}(2017)\citenamefont{Seiferle, von~der
  Wense, and Thirolf}}]{Seiferle_PRL_2017}
\bibinfo{author}{\bibfnamefont{B.}~\bibnamefont{Seiferle}},
  \bibinfo{author}{\bibfnamefont{L.}~\bibnamefont{von~der Wense}},
  \bibnamefont{and} \bibinfo{author}{\bibfnamefont{P.~G.}
  \bibnamefont{Thirolf}}, \bibinfo{journal}{Phys. Rev. Lett.}
  \textbf{\bibinfo{volume}{118}}, \bibinfo{pages}{042501}
  (\bibinfo{year}{2017}),
  \urlprefix\url{https://link.aps.org/doi/10.1103/PhysRevLett.118.042501}.

\bibitem[{\citenamefont{Elwell et~al.}(2024)\citenamefont{Elwell, Schneider,
  Jeet, Terhune, Morgan, Alexandrova, Tran~Tan, Derevianko, and
  Hudson}}]{elwell2024laser}
\bibinfo{author}{\bibfnamefont{R.}~\bibnamefont{Elwell}},
  \bibinfo{author}{\bibfnamefont{C.}~\bibnamefont{Schneider}},
  \bibinfo{author}{\bibfnamefont{J.}~\bibnamefont{Jeet}},
  \bibinfo{author}{\bibfnamefont{J.~E.~S.} \bibnamefont{Terhune}},
  \bibinfo{author}{\bibfnamefont{H.~W.~T.} \bibnamefont{Morgan}},
  \bibinfo{author}{\bibfnamefont{A.~N.} \bibnamefont{Alexandrova}},
  \bibinfo{author}{\bibfnamefont{H.~B.} \bibnamefont{Tran~Tan}},
  \bibinfo{author}{\bibfnamefont{A.}~\bibnamefont{Derevianko}},
  \bibnamefont{and} \bibinfo{author}{\bibfnamefont{E.~R.}
  \bibnamefont{Hudson}}, \bibinfo{journal}{Phys. Rev. Lett.}
  \textbf{\bibinfo{volume}{133}}, \bibinfo{pages}{013201}
  (\bibinfo{year}{2024}),
  \urlprefix\url{https://link.aps.org/doi/10.1103/PhysRevLett.133.013201}.

\bibitem[{\citenamefont{Zhang et~al.}(2024)\citenamefont{Zhang, Ooi, Higgins,
  Doyle, von~der Wense, Beeks, Leitner, Kazakov, Li, Thirolf
  et~al.}}]{JunYe2024}
\bibinfo{author}{\bibfnamefont{C.}~\bibnamefont{Zhang}},
  \bibinfo{author}{\bibfnamefont{T.}~\bibnamefont{Ooi}},
  \bibinfo{author}{\bibfnamefont{J.~S.} \bibnamefont{Higgins}},
  \bibinfo{author}{\bibfnamefont{J.~F.} \bibnamefont{Doyle}},
  \bibinfo{author}{\bibfnamefont{L.}~\bibnamefont{von~der Wense}},
  \bibinfo{author}{\bibfnamefont{K.}~\bibnamefont{Beeks}},
  \bibinfo{author}{\bibfnamefont{A.}~\bibnamefont{Leitner}},
  \bibinfo{author}{\bibfnamefont{G.}~\bibnamefont{Kazakov}},
  \bibinfo{author}{\bibfnamefont{P.}~\bibnamefont{Li}},
  \bibinfo{author}{\bibfnamefont{P.~G.} \bibnamefont{Thirolf}},
  \bibnamefont{et~al.} (\bibinfo{year}{2024}),
  \bibinfo{note}{arXiv:2406.18719v1 [physics.atom-ph]}.

\bibitem[{\citenamefont{Dykhne and Tkalya}(1998)}]{Dyk98}
\bibinfo{author}{\bibfnamefont{A.~M.} \bibnamefont{Dykhne}} \bibnamefont{and}
  \bibinfo{author}{\bibfnamefont{E.~V.} \bibnamefont{Tkalya}},
  \bibinfo{journal}{JETP Lett.} \textbf{\bibinfo{volume}{67}},
  \bibinfo{pages}{251} (\bibinfo{year}{1998}).

\bibitem[{\citenamefont{Tkalya et~al.}(2015)\citenamefont{Tkalya, Schneider,
  Jeet, and Hudson}}]{Tkalya15}
\bibinfo{author}{\bibfnamefont{E.~V.} \bibnamefont{Tkalya}},
  \bibinfo{author}{\bibfnamefont{C.}~\bibnamefont{Schneider}},
  \bibinfo{author}{\bibfnamefont{J.}~\bibnamefont{Jeet}}, \bibnamefont{and}
  \bibinfo{author}{\bibfnamefont{E.~R.} \bibnamefont{Hudson}},
  \bibinfo{journal}{Phys. Rev. C} \textbf{\bibinfo{volume}{92}},
  \bibinfo{pages}{054324} (\bibinfo{year}{2015}),
  \urlprefix\url{https://link.aps.org/doi/10.1103/PhysRevC.92.054324}.

\bibitem[{\citenamefont{Alaga et~al.}(1955)\citenamefont{Alaga, Alder, Bohr,
  and Mottelson}}]{Alaga55}
\bibinfo{author}{\bibfnamefont{G.}~\bibnamefont{Alaga}},
  \bibinfo{author}{\bibfnamefont{K.}~\bibnamefont{Alder}},
  \bibinfo{author}{\bibfnamefont{A.}~\bibnamefont{Bohr}}, \bibnamefont{and}
  \bibinfo{author}{\bibfnamefont{B.}~\bibnamefont{Mottelson}},
  \bibinfo{journal}{Kgl. Danske Videnskab. Selskab Mat.-fys. Medd.}
  \textbf{\bibinfo{volume}{29 \textmd{(No. 9)}}}, \bibinfo{pages}{1}
  (\bibinfo{year}{1955}).

\bibitem[{\citenamefont{Nilsson}(1955)}]{Nilsson1955}
\bibinfo{author}{\bibfnamefont{S.~G.} \bibnamefont{Nilsson}},
  \bibinfo{journal}{Kgl. Danske Videnskab. Selskab Mat.-fys. Medd.}
  \textbf{\bibinfo{volume}{29 \textmd{(No. 16)}}}, \bibinfo{pages}{1}
  (\bibinfo{year}{1955}).

\bibitem[{\citenamefont{Gulda et~al.}(2002)\citenamefont{Gulda, Kurcewicz, Aas,
  Borge, Burke, Fogelberg, Grant, Hagebø, Kaffrell, Kvasil et~al.}}]{Gulda02}
\bibinfo{author}{\bibfnamefont{K.}~\bibnamefont{Gulda}},
  \bibinfo{author}{\bibfnamefont{W.}~\bibnamefont{Kurcewicz}},
  \bibinfo{author}{\bibfnamefont{A.}~\bibnamefont{Aas}},
  \bibinfo{author}{\bibfnamefont{M.}~\bibnamefont{Borge}},
  \bibinfo{author}{\bibfnamefont{D.}~\bibnamefont{Burke}},
  \bibinfo{author}{\bibfnamefont{B.}~\bibnamefont{Fogelberg}},
  \bibinfo{author}{\bibfnamefont{I.}~\bibnamefont{Grant}},
  \bibinfo{author}{\bibfnamefont{E.}~\bibnamefont{Hagebø}},
  \bibinfo{author}{\bibfnamefont{N.}~\bibnamefont{Kaffrell}},
  \bibinfo{author}{\bibfnamefont{J.}~\bibnamefont{Kvasil}},
  \bibnamefont{et~al.}, \bibinfo{journal}{Nuclear Physics A}
  \textbf{\bibinfo{volume}{703}}, \bibinfo{pages}{45 } (\bibinfo{year}{2002}),
  ISSN \bibinfo{issn}{0375-9474},
  \urlprefix\url{http://www.sciencedirect.com/science/article/pii/S0375947401014567}.

\bibitem[{\citenamefont{Ruchowska et~al.}(2006)\citenamefont{Ruchowska,
  P\l{}\'ociennik, \ifmmode~\dot{Z}\else \.{Z}\fi{}ylicz, Mach, Kvasil, Algora,
  Amzal, B\"ack, Borge, Boutami et~al.}}]{Ruch06}
\bibinfo{author}{\bibfnamefont{E.}~\bibnamefont{Ruchowska}},
  \bibinfo{author}{\bibfnamefont{W.~A.} \bibnamefont{P\l{}\'ociennik}},
  \bibinfo{author}{\bibfnamefont{J.}~\bibnamefont{\ifmmode~\dot{Z}\else
  \.{Z}\fi{}ylicz}}, \bibinfo{author}{\bibfnamefont{H.}~\bibnamefont{Mach}},
  \bibinfo{author}{\bibfnamefont{J.}~\bibnamefont{Kvasil}},
  \bibinfo{author}{\bibfnamefont{A.}~\bibnamefont{Algora}},
  \bibinfo{author}{\bibfnamefont{N.}~\bibnamefont{Amzal}},
  \bibinfo{author}{\bibfnamefont{T.}~\bibnamefont{B\"ack}},
  \bibinfo{author}{\bibfnamefont{M.~G.} \bibnamefont{Borge}},
  \bibinfo{author}{\bibfnamefont{R.}~\bibnamefont{Boutami}},
  \bibnamefont{et~al.}, \bibinfo{journal}{Phys. Rev. C}
  \textbf{\bibinfo{volume}{73}}, \bibinfo{pages}{044326}
  (\bibinfo{year}{2006}).

\bibitem[{\citenamefont{Soloviev}(1976)}]{Sol76}
\bibinfo{author}{\bibfnamefont{V.~G.} \bibnamefont{Soloviev}},
  \emph{\bibinfo{title}{Theory of Complex Nuclei}}
  (\bibinfo{publisher}{Pergamon Press, Oxford}, \bibinfo{year}{1976}).

\bibitem[{\citenamefont{He and Ren}(2007)}]{JPG07_He}
\bibinfo{author}{\bibfnamefont{X.-T.} \bibnamefont{He}} \bibnamefont{and}
  \bibinfo{author}{\bibfnamefont{Z.-Z.} \bibnamefont{Ren}},
  \bibinfo{journal}{J. Phys. G: Nucl. Part. Phys.}
  \textbf{\bibinfo{volume}{34}}, \bibinfo{pages}{1611} (\bibinfo{year}{2007}).

\bibitem[{\citenamefont{He and Ren}(2008)}]{NPA08_He}
\bibinfo{author}{\bibfnamefont{X.-T.} \bibnamefont{He}} \bibnamefont{and}
  \bibinfo{author}{\bibfnamefont{Z.-Z.} \bibnamefont{Ren}},
  \bibinfo{journal}{Nucl. Phys. A} \textbf{\bibinfo{volume}{806}},
  \bibinfo{pages}{117} (\bibinfo{year}{2008}).

\bibitem[{\citenamefont{Litvinova et~al.}(2009)\citenamefont{Litvinova,
  Feldmeier, Dobaczewski, and Flambaum}}]{PRC09_Litvinova_HFB}
\bibinfo{author}{\bibfnamefont{E.}~\bibnamefont{Litvinova}},
  \bibinfo{author}{\bibfnamefont{H.}~\bibnamefont{Feldmeier}},
  \bibinfo{author}{\bibfnamefont{J.}~\bibnamefont{Dobaczewski}},
  \bibnamefont{and} \bibinfo{author}{\bibfnamefont{V.}~\bibnamefont{Flambaum}},
  \bibinfo{journal}{Phys. Rev. C} \textbf{\bibinfo{volume}{79}},
  \bibinfo{pages}{064303} (\bibinfo{year}{2009}).

\bibitem[{\citenamefont{Minkov and P\'alffy}(2017)}]{Minkov_Palffy_PRL_2017}
\bibinfo{author}{\bibfnamefont{N.}~\bibnamefont{Minkov}} \bibnamefont{and}
  \bibinfo{author}{\bibfnamefont{A.}~\bibnamefont{P\'alffy}},
  \bibinfo{journal}{Phys. Rev. Lett.} \textbf{\bibinfo{volume}{118}},
  \bibinfo{pages}{212501} (\bibinfo{year}{2017}).

\bibitem[{\citenamefont{Minkov and P\'alffy}(2019)}]{Minkov_Palffy_PRL_2019}
\bibinfo{author}{\bibfnamefont{N.}~\bibnamefont{Minkov}} \bibnamefont{and}
  \bibinfo{author}{\bibfnamefont{A.}~\bibnamefont{P\'alffy}},
  \bibinfo{journal}{Phys. Rev. Lett.} \textbf{\bibinfo{volume}{122}},
  \bibinfo{pages}{162502} (\bibinfo{year}{2019}).

\bibitem[{\citenamefont{Minkov and P\'alffy}(2021)}]{Minkov_Palffy_PRC_2021}
\bibinfo{author}{\bibfnamefont{N.}~\bibnamefont{Minkov}} \bibnamefont{and}
  \bibinfo{author}{\bibfnamefont{A.}~\bibnamefont{P\'alffy}},
  \bibinfo{journal}{Phys. Rev. C} \textbf{\bibinfo{volume}{103}},
  \bibinfo{pages}{014313} (\bibinfo{year}{2021}).

\bibitem[{\citenamefont{Nilsson and Ragnarsson}(1995)}]{NR1995}
\bibinfo{author}{\bibfnamefont{S.~G.} \bibnamefont{Nilsson}} \bibnamefont{and}
  \bibinfo{author}{\bibfnamefont{I.}~\bibnamefont{Ragnarsson}},
  \emph{\bibinfo{title}{Shapes and Shells in Nuclear Structure}}
  (\bibinfo{publisher}{Cambridge University Press, Cambridge},
  \bibinfo{year}{1995}).

\bibitem[{\citenamefont{Minkov et~al.}(2006)\citenamefont{Minkov, Yotov,
  Drenska, Scheid, Bonatsos, Lenis, and Petrellis}}]{b2b3mod}
\bibinfo{author}{\bibfnamefont{N.}~\bibnamefont{Minkov}},
  \bibinfo{author}{\bibfnamefont{P.}~\bibnamefont{Yotov}},
  \bibinfo{author}{\bibfnamefont{S.}~\bibnamefont{Drenska}},
  \bibinfo{author}{\bibfnamefont{W.}~\bibnamefont{Scheid}},
  \bibinfo{author}{\bibfnamefont{D.}~\bibnamefont{Bonatsos}},
  \bibinfo{author}{\bibfnamefont{D.}~\bibnamefont{Lenis}}, \bibnamefont{and}
  \bibinfo{author}{\bibfnamefont{D.}~\bibnamefont{Petrellis}},
  \bibinfo{journal}{Phys. Rev. C} \textbf{\bibinfo{volume}{73}},
  \bibinfo{pages}{044315} (\bibinfo{year}{2006}),
  \urlprefix\url{https://link.aps.org/doi/10.1103/PhysRevC.73.044315}.

\bibitem[{\citenamefont{Minkov et~al.}(2007)\citenamefont{Minkov, Drenska,
  Yotov, Lalkovski, Bonatsos, and Scheid}}]{b2b3odd}
\bibinfo{author}{\bibfnamefont{N.}~\bibnamefont{Minkov}},
  \bibinfo{author}{\bibfnamefont{S.}~\bibnamefont{Drenska}},
  \bibinfo{author}{\bibfnamefont{P.}~\bibnamefont{Yotov}},
  \bibinfo{author}{\bibfnamefont{S.}~\bibnamefont{Lalkovski}},
  \bibinfo{author}{\bibfnamefont{D.}~\bibnamefont{Bonatsos}}, \bibnamefont{and}
  \bibinfo{author}{\bibfnamefont{W.}~\bibnamefont{Scheid}},
  \bibinfo{journal}{Phys. Rev. C} \textbf{\bibinfo{volume}{76}},
  \bibinfo{pages}{034324} (\bibinfo{year}{2007}),
  \urlprefix\url{https://link.aps.org/doi/10.1103/PhysRevC.76.034324}.

\bibitem[{\citenamefont{Minkov}(2013)}]{NM13}
\bibinfo{author}{\bibfnamefont{N.}~\bibnamefont{Minkov}},
  \bibinfo{journal}{Physica Scripta} \textbf{\bibinfo{volume}{T154}},
  \bibinfo{pages}{014017} (\bibinfo{year}{2013}),
  \urlprefix\url{http://stacks.iop.org/1402-4896/2013/i=T154/a=014017}.

\bibitem[{\citenamefont{Minkov et~al.}(2012)\citenamefont{Minkov, Drenska,
  Strecker, Scheid, and Lenske}}]{MDSSL12}
\bibinfo{author}{\bibfnamefont{N.}~\bibnamefont{Minkov}},
  \bibinfo{author}{\bibfnamefont{S.}~\bibnamefont{Drenska}},
  \bibinfo{author}{\bibfnamefont{M.}~\bibnamefont{Strecker}},
  \bibinfo{author}{\bibfnamefont{W.}~\bibnamefont{Scheid}}, \bibnamefont{and}
  \bibinfo{author}{\bibfnamefont{H.}~\bibnamefont{Lenske}},
  \bibinfo{journal}{Phys. Rev. C} \textbf{\bibinfo{volume}{85}},
  \bibinfo{pages}{034306} (\bibinfo{year}{2012}),
  \urlprefix\url{https://link.aps.org/doi/10.1103/PhysRevC.85.034306}.

\bibitem[{\citenamefont{Minkov et~al.}(2013)\citenamefont{Minkov, Drenska,
  Drumev, Strecker, Lenske, and Scheid}}]{MDDSLS13}
\bibinfo{author}{\bibfnamefont{N.}~\bibnamefont{Minkov}},
  \bibinfo{author}{\bibfnamefont{S.}~\bibnamefont{Drenska}},
  \bibinfo{author}{\bibfnamefont{K.}~\bibnamefont{Drumev}},
  \bibinfo{author}{\bibfnamefont{M.}~\bibnamefont{Strecker}},
  \bibinfo{author}{\bibfnamefont{H.}~\bibnamefont{Lenske}}, \bibnamefont{and}
  \bibinfo{author}{\bibfnamefont{W.}~\bibnamefont{Scheid}},
  \bibinfo{journal}{Phys. Rev. C} \textbf{\bibinfo{volume}{88}},
  \bibinfo{pages}{064310} (\bibinfo{year}{2013}),
  \urlprefix\url{https://link.aps.org/doi/10.1103/PhysRevC.88.064310}.

\bibitem[{\citenamefont{Cwiok et~al.}(1987)\citenamefont{Cwiok, Dudek,
  Nazarewicz, Skalski, and Werner}}]{qocsmod}
\bibinfo{author}{\bibfnamefont{S.}~\bibnamefont{Cwiok}},
  \bibinfo{author}{\bibfnamefont{J.}~\bibnamefont{Dudek}},
  \bibinfo{author}{\bibfnamefont{W.}~\bibnamefont{Nazarewicz}},
  \bibinfo{author}{\bibfnamefont{J.}~\bibnamefont{Skalski}}, \bibnamefont{and}
  \bibinfo{author}{\bibfnamefont{T.}~\bibnamefont{Werner}},
  \bibinfo{journal}{Computer Physics Communications}
  \textbf{\bibinfo{volume}{46}}, \bibinfo{pages}{379 } (\bibinfo{year}{1987}),
  ISSN \bibinfo{issn}{0010-4655},
  \urlprefix\url{http://www.sciencedirect.com/science/article/pii/0010465587900932}.

\bibitem[{\citenamefont{Bonneau et~al.}(2015)\citenamefont{Bonneau, Minkov,
  Duc, Quentin, and Bartel}}]{PRC15_hfbcs}
\bibinfo{author}{\bibfnamefont{L.}~\bibnamefont{Bonneau}},
  \bibinfo{author}{\bibfnamefont{N.}~\bibnamefont{Minkov}},
  \bibinfo{author}{\bibfnamefont{D.~D.} \bibnamefont{Duc}},
  \bibinfo{author}{\bibfnamefont{P.}~\bibnamefont{Quentin}}, \bibnamefont{and}
  \bibinfo{author}{\bibfnamefont{J.}~\bibnamefont{Bartel}},
  \bibinfo{journal}{Phys. Rev. C} \textbf{\bibinfo{volume}{91}},
  \bibinfo{pages}{054307} (\bibinfo{year}{2015}).

\bibitem[{\citenamefont{Beiner et~al.}(1975)\citenamefont{Beiner, Flocard,
  Giai, and Quentin}}]{Beiner75_SIII}
\bibinfo{author}{\bibfnamefont{M.}~\bibnamefont{Beiner}},
  \bibinfo{author}{\bibfnamefont{H.}~\bibnamefont{Flocard}},
  \bibinfo{author}{\bibfnamefont{N.~V.} \bibnamefont{Giai}}, \bibnamefont{and}
  \bibinfo{author}{\bibfnamefont{P.}~\bibnamefont{Quentin}},
  \bibinfo{journal}{Nucl. Phys. A} \textbf{\bibinfo{volume}{238}},
  \bibinfo{pages}{29} (\bibinfo{year}{1975}).

\bibitem[{\citenamefont{Vautherin}(1973)}]{Vautherin73}
\bibinfo{author}{\bibfnamefont{D.}~\bibnamefont{Vautherin}},
  \bibinfo{journal}{Phys. Rev. C} \textbf{\bibinfo{volume}{7}},
  \bibinfo{pages}{296} (\bibinfo{year}{1973}).

\bibitem[{\citenamefont{Bonneau et~al.}(2019)\citenamefont{Bonneau, Quentin,
  Minkov, Ivanova, Bartel, Molique, and Koh}}]{BJP19_hfbcs_isomers}
\bibinfo{author}{\bibfnamefont{L.}~\bibnamefont{Bonneau}},
  \bibinfo{author}{\bibfnamefont{P.}~\bibnamefont{Quentin}},
  \bibinfo{author}{\bibfnamefont{N.}~\bibnamefont{Minkov}},
  \bibinfo{author}{\bibfnamefont{D.}~\bibnamefont{Ivanova}},
  \bibinfo{author}{\bibfnamefont{J.}~\bibnamefont{Bartel}},
  \bibinfo{author}{\bibfnamefont{H.}~\bibnamefont{Molique}}, \bibnamefont{and}
  \bibinfo{author}{\bibfnamefont{M.-H.} \bibnamefont{Koh}},
  \bibinfo{journal}{Bulg. J. Phys.} \textbf{\bibinfo{volume}{46}},
  \bibinfo{pages}{366} (\bibinfo{year}{2019}).

\bibitem[{\citenamefont{Quentin et~al.}(2021)\citenamefont{Quentin, Bonneau,
  Minkov, Ivanova, Bartel, Molique, and Koh}}]{BJP21_hfbcs_isomers}
\bibinfo{author}{\bibfnamefont{P.}~\bibnamefont{Quentin}},
  \bibinfo{author}{\bibfnamefont{L.}~\bibnamefont{Bonneau}},
  \bibinfo{author}{\bibfnamefont{N.}~\bibnamefont{Minkov}},
  \bibinfo{author}{\bibfnamefont{D.}~\bibnamefont{Ivanova}},
  \bibinfo{author}{\bibfnamefont{J.}~\bibnamefont{Bartel}},
  \bibinfo{author}{\bibfnamefont{H.}~\bibnamefont{Molique}}, \bibnamefont{and}
  \bibinfo{author}{\bibfnamefont{M.-H.} \bibnamefont{Koh}},
  \bibinfo{journal}{Bulg. J. Phys.} \textbf{\bibinfo{volume}{48}},
  \bibinfo{pages}{634} (\bibinfo{year}{2021}).

\bibitem[{\citenamefont{Minkov et~al.}(2022)\citenamefont{Minkov, Bonneau,
  Quentin, Bartel, Molique, and Ivanova}}]{PRC22_hfbcs_isomers}
\bibinfo{author}{\bibfnamefont{N.}~\bibnamefont{Minkov}},
  \bibinfo{author}{\bibfnamefont{L.}~\bibnamefont{Bonneau}},
  \bibinfo{author}{\bibfnamefont{P.}~\bibnamefont{Quentin}},
  \bibinfo{author}{\bibfnamefont{J.}~\bibnamefont{Bartel}},
  \bibinfo{author}{\bibfnamefont{H.}~\bibnamefont{Molique}}, \bibnamefont{and}
  \bibinfo{author}{\bibfnamefont{D.}~\bibnamefont{Ivanova}},
  \bibinfo{journal}{Phys. Rev. C} \textbf{\bibinfo{volume}{105}},
  \bibinfo{pages}{044329} (\bibinfo{year}{2022}).

\bibitem[{\citenamefont{Minkov et~al.}(2024)\citenamefont{Minkov, Bonneau,
  Quentin, Bartel, Molique, and Koh}}]{PRC24_hfbcs_isomers}
\bibinfo{author}{\bibfnamefont{N.}~\bibnamefont{Minkov}},
  \bibinfo{author}{\bibfnamefont{L.}~\bibnamefont{Bonneau}},
  \bibinfo{author}{\bibfnamefont{P.}~\bibnamefont{Quentin}},
  \bibinfo{author}{\bibfnamefont{J.}~\bibnamefont{Bartel}},
  \bibinfo{author}{\bibfnamefont{H.}~\bibnamefont{Molique}}, \bibnamefont{and}
  \bibinfo{author}{\bibfnamefont{M.-H.} \bibnamefont{Koh}},
  \bibinfo{journal}{Phys. Rev. C} \textbf{\bibinfo{volume}{109}},
  \bibinfo{pages}{064315} (\bibinfo{year}{2024}),
  \urlprefix\url{https://doi.org/10.1103/PhysRevC.109.064315}.

\bibitem[{\citenamefont{Dobaczewski et~al.}(2021)\citenamefont{Dobaczewski,
  Baczyk, Becker, Bender, Bennaceur, Bonnard, Gao, Idini, Konieczka,
  Kortelainen et~al.}}]{JPG21_Dobaczewski}
\bibinfo{author}{\bibfnamefont{J.}~\bibnamefont{Dobaczewski}},
  \bibinfo{author}{\bibfnamefont{P.}~\bibnamefont{Baczyk}},
  \bibinfo{author}{\bibfnamefont{P.}~\bibnamefont{Becker}},
  \bibinfo{author}{\bibfnamefont{M.}~\bibnamefont{Bender}},
  \bibinfo{author}{\bibfnamefont{K.}~\bibnamefont{Bennaceur}},
  \bibinfo{author}{\bibfnamefont{J.}~\bibnamefont{Bonnard}},
  \bibinfo{author}{\bibfnamefont{Y.}~\bibnamefont{Gao}},
  \bibinfo{author}{\bibfnamefont{A.}~\bibnamefont{Idini}},
  \bibinfo{author}{\bibfnamefont{M.}~\bibnamefont{Konieczka}},
  \bibinfo{author}{\bibfnamefont{M.}~\bibnamefont{Kortelainen}},
  \bibnamefont{et~al.}, \bibinfo{journal}{J. Phys. G: Nucl. Part. Phys.}
  \textbf{\bibinfo{volume}{48}}, \bibinfo{pages}{102001}
  (\bibinfo{year}{2021}).

\bibitem[{\citenamefont{Ryssens et~al.}(2015)\citenamefont{Ryssens, Hellemans,
  Bender, and Heenen}}]{cpc15_Ryssens}
\bibinfo{author}{\bibfnamefont{W.}~\bibnamefont{Ryssens}},
  \bibinfo{author}{\bibfnamefont{V.}~\bibnamefont{Hellemans}},
  \bibinfo{author}{\bibfnamefont{M.}~\bibnamefont{Bender}}, \bibnamefont{and}
  \bibinfo{author}{\bibfnamefont{P.~H.} \bibnamefont{Heenen}},
  \bibinfo{journal}{Comp. Phys. Comm.} \textbf{\bibinfo{volume}{187}},
  \bibinfo{pages}{175} (\bibinfo{year}{2015}).

\bibitem[{\citenamefont{Chen et~al.}(2022)\citenamefont{Chen, Lia, Schuetrumpf,
  Reinhard, and Nazarewicz}}]{cpc22_Sky3D}
\bibinfo{author}{\bibfnamefont{M.}~\bibnamefont{Chen}},
  \bibinfo{author}{\bibfnamefont{T.}~\bibnamefont{Lia}},
  \bibinfo{author}{\bibfnamefont{B.}~\bibnamefont{Schuetrumpf}},
  \bibinfo{author}{\bibfnamefont{P.~G.} \bibnamefont{Reinhard}},
  \bibnamefont{and}
  \bibinfo{author}{\bibfnamefont{W.}~\bibnamefont{Nazarewicz}},
  \bibinfo{journal}{Comp. Phys. Comm.} \textbf{\bibinfo{volume}{276}},
  \bibinfo{pages}{108344} (\bibinfo{year}{2022}).

\bibitem[{\citenamefont{Flocard et~al.}(1973)\citenamefont{Flocard, Quentin,
  Kerman, and Vautherin}}]{Floc73}
\bibinfo{author}{\bibfnamefont{H.}~\bibnamefont{Flocard}},
  \bibinfo{author}{\bibfnamefont{P.}~\bibnamefont{Quentin}},
  \bibinfo{author}{\bibfnamefont{A.~K.} \bibnamefont{Kerman}},
  \bibnamefont{and}
  \bibinfo{author}{\bibfnamefont{D.}~\bibnamefont{Vautherin}},
  \bibinfo{journal}{Nucl. Phys. A} \textbf{\bibinfo{volume}{203}},
  \bibinfo{pages}{433} (\bibinfo{year}{1973}).

\bibitem[{\citenamefont{Pillet et~al.}(2002)\citenamefont{Pillet, Quentin, and
  Libert}}]{Pillet2002}
\bibinfo{author}{\bibfnamefont{N.}~\bibnamefont{Pillet}},
  \bibinfo{author}{\bibfnamefont{P.}~\bibnamefont{Quentin}}, \bibnamefont{and}
  \bibinfo{author}{\bibfnamefont{J.}~\bibnamefont{Libert}},
  \bibinfo{journal}{Nucl. Phys. A} \textbf{\bibinfo{volume}{697}},
  \bibinfo{pages}{141} (\bibinfo{year}{2002}).

\bibitem[{\citenamefont{Bonche et~al.}(1985)\citenamefont{Bonche, Flocard,
  Heenen, Krieger, and Weiss}}]{Bonche1985}
\bibinfo{author}{\bibfnamefont{P.}~\bibnamefont{Bonche}},
  \bibinfo{author}{\bibfnamefont{H.}~\bibnamefont{Flocard}},
  \bibinfo{author}{\bibfnamefont{P.-H.} \bibnamefont{Heenen}},
  \bibinfo{author}{\bibfnamefont{S.~J.} \bibnamefont{Krieger}},
  \bibnamefont{and} \bibinfo{author}{\bibfnamefont{M.~S.} \bibnamefont{Weiss}},
  \bibinfo{journal}{Nucl. Phys. A} \textbf{\bibinfo{volume}{443}},
  \bibinfo{pages}{39} (\bibinfo{year}{1985}).

\bibitem[{\citenamefont{Nor et~al.}(2019)\citenamefont{Nor, Rezle, Kelvin-Lee,
  Koh, Bonneau, and Quentin}}]{PRC19GnGpfit}
\bibinfo{author}{\bibfnamefont{N.~M.} \bibnamefont{Nor}},
  \bibinfo{author}{\bibfnamefont{N.-A.} \bibnamefont{Rezle}},
  \bibinfo{author}{\bibfnamefont{K.-W.} \bibnamefont{Kelvin-Lee}},
  \bibinfo{author}{\bibfnamefont{M.-H.} \bibnamefont{Koh}},
  \bibinfo{author}{\bibfnamefont{L.}~\bibnamefont{Bonneau}}, \bibnamefont{and}
  \bibinfo{author}{\bibfnamefont{P.}~\bibnamefont{Quentin}},
  \bibinfo{journal}{Phys. Rev. C} \textbf{\bibinfo{volume}{99}},
  \bibinfo{pages}{064306} (\bibinfo{year}{2019}).

\bibitem[{\citenamefont{Inglis}(1954)}]{Inglis54}
\bibinfo{author}{\bibfnamefont{D.~R.} \bibnamefont{Inglis}},
  \bibinfo{journal}{Phys. Rev.} \textbf{\bibinfo{volume}{96}},
  \bibinfo{pages}{1059} (\bibinfo{year}{1954}),
  \urlprefix\url{https://link.aps.org/doi/10.1103/PhysRev.96.1059}.

\bibitem[{\citenamefont{Inglis}(1955)}]{Inglis55}
\bibinfo{author}{\bibfnamefont{D.~R.} \bibnamefont{Inglis}},
  \bibinfo{journal}{Phys. Rev.} \textbf{\bibinfo{volume}{97}},
  \bibinfo{pages}{701} (\bibinfo{year}{1955}),
  \urlprefix\url{https://link.aps.org/doi/10.1103/PhysRev.97.701}.

\bibitem[{\citenamefont{Belyaev}(1961)}]{Belyaev61}
\bibinfo{author}{\bibfnamefont{S.~T.} \bibnamefont{Belyaev}},
  \bibinfo{journal}{Nucl. Phys.} \textbf{\bibinfo{volume}{24}},
  \bibinfo{pages}{322} (\bibinfo{year}{1961}).

\bibitem[{\citenamefont{Yuldashbaeva et~al.}(1999)\citenamefont{Yuldashbaeva,
  Libert, Quentin, and Girod}}]{Yuldashbaeva1999}
\bibinfo{author}{\bibfnamefont{E.~K.} \bibnamefont{Yuldashbaeva}},
  \bibinfo{author}{\bibfnamefont{J.}~\bibnamefont{Libert}},
  \bibinfo{author}{\bibfnamefont{P.}~\bibnamefont{Quentin}}, \bibnamefont{and}
  \bibinfo{author}{\bibfnamefont{M.}~\bibnamefont{Girod}},
  \bibinfo{journal}{Phys. Lett. B} \textbf{\bibinfo{volume}{461}},
  \bibinfo{pages}{1} (\bibinfo{year}{1999}).

\bibitem[{\citenamefont{Libert et~al.}(1999)\citenamefont{Libert, Girod, and
  Delaroche}}]{Libert99}
\bibinfo{author}{\bibfnamefont{J.}~\bibnamefont{Libert}},
  \bibinfo{author}{\bibfnamefont{M.}~\bibnamefont{Girod}}, \bibnamefont{and}
  \bibinfo{author}{\bibfnamefont{J.~P.} \bibnamefont{Delaroche}},
  \bibinfo{journal}{Phys. Rev. C} \textbf{\bibinfo{volume}{60}},
  \bibinfo{pages}{054301} (\bibinfo{year}{1999}).

\bibitem[{ens()}]{ensdf}
\bibinfo{howpublished}{\url{http://www.nndc.bnl.gov/ensdf/}}.

\bibitem[{\citenamefont{Nomura et~al.}(2014)\citenamefont{Nomura, Vretenar,
  Nik\v{s}i\'{c}, and Lu}}]{Nomura2014}
\bibinfo{author}{\bibfnamefont{K.}~\bibnamefont{Nomura}},
  \bibinfo{author}{\bibfnamefont{D.}~\bibnamefont{Vretenar}},
  \bibinfo{author}{\bibfnamefont{T.}~\bibnamefont{Nik\v{s}i\'{c}}},
  \bibnamefont{and} \bibinfo{author}{\bibfnamefont{B.-N.} \bibnamefont{Lu}},
  \bibinfo{journal}{Phys. Rev. C} \textbf{\bibinfo{volume}{89}},
  \bibinfo{pages}{024312} (\bibinfo{year}{2014}),
  \urlprefix\url{https://link.aps.org/doi/10.1103/PhysRevC.89.024312}.

\bibitem[{\citenamefont{Gustafsson et~al.}(1971)\citenamefont{Gustafsson,
  M\"oller, and Nilsson}}]{Gustafsson71}
\bibinfo{author}{\bibfnamefont{C.}~\bibnamefont{Gustafsson}},
  \bibinfo{author}{\bibfnamefont{P.}~\bibnamefont{M\"oller}}, \bibnamefont{and}
  \bibinfo{author}{\bibfnamefont{S.~G.} \bibnamefont{Nilsson}},
  \bibinfo{journal}{Phys. Lett. B} \textbf{\bibinfo{volume}{34}},
  \bibinfo{pages}{349} (\bibinfo{year}{1971}).

\bibitem[{\citenamefont{Johansson}(1961)}]{Johansson61}
\bibinfo{author}{\bibfnamefont{S.~A.~E.} \bibnamefont{Johansson}},
  \bibinfo{journal}{Nucl. Phys.} \textbf{\bibinfo{volume}{22}},
  \bibinfo{pages}{529} (\bibinfo{year}{1961}).

\bibitem[{\citenamefont{Ban et~al.}(2010)\citenamefont{Ban, Dobaczewski, Engel,
  and Shukla}}]{Ban10}
\bibinfo{author}{\bibfnamefont{S.}~\bibnamefont{Ban}},
  \bibinfo{author}{\bibfnamefont{J.}~\bibnamefont{Dobaczewski}},
  \bibinfo{author}{\bibfnamefont{J.}~\bibnamefont{Engel}}, \bibnamefont{and}
  \bibinfo{author}{\bibfnamefont{A.}~\bibnamefont{Shukla}},
  \bibinfo{journal}{Phys. Rev. C} \textbf{\bibinfo{volume}{82}},
  \bibinfo{pages}{015501} (\bibinfo{year}{2010}).

\bibitem[{\citenamefont{Afanasjev and Abdurazakov}(2013)}]{Afanasjev13}
\bibinfo{author}{\bibfnamefont{A.~V.} \bibnamefont{Afanasjev}}
  \bibnamefont{and}
  \bibinfo{author}{\bibfnamefont{O.}~\bibnamefont{Abdurazakov}},
  \bibinfo{journal}{Phys. Rev. C} \textbf{\bibinfo{volume}{88}},
  \bibinfo{pages}{014320} (\bibinfo{year}{2013}).

\bibitem[{\citenamefont{Safronova et~al.}(2013)\citenamefont{Safronova,
  Safronova, Radnaev, Campbell, and Kuzmich}}]{Safronova13}
\bibinfo{author}{\bibfnamefont{M.~S.} \bibnamefont{Safronova}},
  \bibinfo{author}{\bibfnamefont{U.~I.} \bibnamefont{Safronova}},
  \bibinfo{author}{\bibfnamefont{A.~G.} \bibnamefont{Radnaev}},
  \bibinfo{author}{\bibfnamefont{C.~J.} \bibnamefont{Campbell}},
  \bibnamefont{and} \bibinfo{author}{\bibfnamefont{A.}~\bibnamefont{Kuzmich}},
  \bibinfo{journal}{Phys. Rev. A} \textbf{\bibinfo{volume}{88}},
  \bibinfo{pages}{060501(R)} (\bibinfo{year}{2013}),
  \urlprefix\url{https://link.aps.org/doi/10.1103/PhysRevA.88.060501}.

\bibitem[{\citenamefont{Gerstenkorn et~al.}(1974)\citenamefont{Gerstenkorn,
  Luc, Verges, Englekemeir, Gindler, and Tomkins}}]{Gerstenkorn74}
\bibinfo{author}{\bibfnamefont{S.}~\bibnamefont{Gerstenkorn}},
  \bibinfo{author}{\bibfnamefont{P.}~\bibnamefont{Luc}},
  \bibinfo{author}{\bibfnamefont{J.}~\bibnamefont{Verges}},
  \bibinfo{author}{\bibfnamefont{D.~W.} \bibnamefont{Englekemeir}},
  \bibinfo{author}{\bibfnamefont{J.~E.} \bibnamefont{Gindler}},
  \bibnamefont{and} \bibinfo{author}{\bibfnamefont{F.~S.}
  \bibnamefont{Tomkins}}, \bibinfo{journal}{J. Phys. (Paris)}
  \textbf{\bibinfo{volume}{35}}, \bibinfo{pages}{483} (\bibinfo{year}{1974}).

\bibitem[{\citenamefont{Thielking et~al.}(2018)\citenamefont{Thielking,
  Okhapkin, Glowacki, Meier, von~der Wense, Seiferle, D{\"u}llmann, Thirolf,
  and Peik}}]{Thielking2018}
\bibinfo{author}{\bibfnamefont{J.}~\bibnamefont{Thielking}},
  \bibinfo{author}{\bibfnamefont{M.~V.} \bibnamefont{Okhapkin}},
  \bibinfo{author}{\bibfnamefont{P.}~\bibnamefont{Glowacki}},
  \bibinfo{author}{\bibfnamefont{D.~M.} \bibnamefont{Meier}},
  \bibinfo{author}{\bibfnamefont{L.}~\bibnamefont{von~der Wense}},
  \bibinfo{author}{\bibfnamefont{B.}~\bibnamefont{Seiferle}},
  \bibinfo{author}{\bibfnamefont{C.~E.} \bibnamefont{D{\"u}llmann}},
  \bibinfo{author}{\bibfnamefont{P.~G.} \bibnamefont{Thirolf}},
  \bibnamefont{and} \bibinfo{author}{\bibfnamefont{P.}~\bibnamefont{Peik}},
  \bibinfo{journal}{Nature (London)} \textbf{\bibinfo{volume}{556}},
  \bibinfo{pages}{321} (\bibinfo{year}{2018}).

\bibitem[{\citenamefont{M\"uller et~al.}(2018)\citenamefont{M\"uller, Maiorova,
  Fritzsche, Volotka, Beerwerth, Glowacki, Thielking, Meier, Okhapkin, Peik
  et~al.}}]{Mueller18}
\bibinfo{author}{\bibfnamefont{R.~A.} \bibnamefont{M\"uller}},
  \bibinfo{author}{\bibfnamefont{A.~V.} \bibnamefont{Maiorova}},
  \bibinfo{author}{\bibfnamefont{S.}~\bibnamefont{Fritzsche}},
  \bibinfo{author}{\bibfnamefont{A.~V.} \bibnamefont{Volotka}},
  \bibinfo{author}{\bibfnamefont{R.}~\bibnamefont{Beerwerth}},
  \bibinfo{author}{\bibfnamefont{P.}~\bibnamefont{Glowacki}},
  \bibinfo{author}{\bibfnamefont{J.}~\bibnamefont{Thielking}},
  \bibinfo{author}{\bibfnamefont{D.-M.} \bibnamefont{Meier}},
  \bibinfo{author}{\bibfnamefont{M.}~\bibnamefont{Okhapkin}},
  \bibinfo{author}{\bibfnamefont{E.}~\bibnamefont{Peik}}, \bibnamefont{et~al.},
  \bibinfo{journal}{Phys. Rev. A} \textbf{\bibinfo{volume}{98}},
  \bibinfo{pages}{020503(R)} (\bibinfo{year}{2018}),
  \urlprefix\url{https://link.aps.org/doi/10.1103/PhysRevA.98.020503}.

\bibitem[{\citenamefont{Hao et~al.}(2012)\citenamefont{Hao, Quentin, and
  Bonneau}}]{Nhan2012}
\bibinfo{author}{\bibfnamefont{T.~V.~N.} \bibnamefont{Hao}},
  \bibinfo{author}{\bibfnamefont{P.}~\bibnamefont{Quentin}}, \bibnamefont{and}
  \bibinfo{author}{\bibfnamefont{L.}~\bibnamefont{Bonneau}},
  \bibinfo{journal}{Phys. Rev. C} \textbf{\bibinfo{volume}{86}},
  \bibinfo{pages}{064307} (\bibinfo{year}{2012}),
  \urlprefix\url{https://link.aps.org/doi/10.1103/PhysRevC.86.064307}.

\bibitem[{\citenamefont{Bonatsos et~al.}(2005)\citenamefont{Bonatsos, Lenis,
  Minkov, Petrellis, and Yotov}}]{Bonat2005}
\bibinfo{author}{\bibfnamefont{D.}~\bibnamefont{Bonatsos}},
  \bibinfo{author}{\bibfnamefont{D.}~\bibnamefont{Lenis}},
  \bibinfo{author}{\bibfnamefont{N.}~\bibnamefont{Minkov}},
  \bibinfo{author}{\bibfnamefont{D.}~\bibnamefont{Petrellis}},
  \bibnamefont{and} \bibinfo{author}{\bibfnamefont{P.}~\bibnamefont{Yotov}},
  \bibinfo{journal}{Phys. Rev. C} \textbf{\bibinfo{volume}{71}},
  \bibinfo{pages}{064309} (\bibinfo{year}{2005}).

\bibitem[{\citenamefont{Nazarewicz et~al.}(1984)\citenamefont{Nazarewicz,
  Olanders, Ragnarsson, J.~Dudek, M\"oller, and Ruchowska}}]{NazOland84}
\bibinfo{author}{\bibfnamefont{W.}~\bibnamefont{Nazarewicz}},
  \bibinfo{author}{\bibfnamefont{P.}~\bibnamefont{Olanders}},
  \bibinfo{author}{\bibfnamefont{I.}~\bibnamefont{Ragnarsson}},
  \bibinfo{author}{\bibfnamefont{G.~A.~L.} \bibnamefont{J.~Dudek}},
  \bibinfo{author}{\bibfnamefont{P.}~\bibnamefont{M\"oller}}, \bibnamefont{and}
  \bibinfo{author}{\bibfnamefont{E.}~\bibnamefont{Ruchowska}},
  \bibinfo{journal}{Nucl. Phys. A} \textbf{\bibinfo{volume}{429}},
  \bibinfo{pages}{269} (\bibinfo{year}{1984}).

\bibitem[{\citenamefont{Roux et~al.}(2024)\citenamefont{Roux, Bark, Lawrie
  et~al.}}]{Roux_EPJA2024}
\bibinfo{author}{\bibfnamefont{D.~G.} \bibnamefont{Roux}},
  \bibinfo{author}{\bibfnamefont{R.~A.} \bibnamefont{Bark}},
  \bibinfo{author}{\bibfnamefont{E.~A.} \bibnamefont{Lawrie}},
  \bibnamefont{et~al.}, \bibinfo{journal}{Eur. Phys. J. A}
  \textbf{\bibinfo{volume}{60}}, \bibinfo{pages}{118} (\bibinfo{year}{2024}),
  \urlprefix\url{https://doi.org/10.1140/epja/s10050-024-01337-z}.

\bibitem[{\citenamefont{Butler and Nazarewicz}(1996)}]{BN96}
\bibinfo{author}{\bibfnamefont{P.~A.} \bibnamefont{Butler}} \bibnamefont{and}
  \bibinfo{author}{\bibfnamefont{W.}~\bibnamefont{Nazarewicz}},
  \bibinfo{journal}{Rev. Mod. Phys.} \textbf{\bibinfo{volume}{68}},
  \bibinfo{pages}{349} (\bibinfo{year}{1996}).

\bibitem[{\citenamefont{Butler}(2016)}]{Butler2016}
\bibinfo{author}{\bibfnamefont{P.~A.} \bibnamefont{Butler}},
  \bibinfo{journal}{J. Phys. G: Nucl. Part. Phys.}
  \textbf{\bibinfo{volume}{43}}, \bibinfo{pages}{073002}
  (\bibinfo{year}{2016}).

\bibitem[{\citenamefont{Stone}(2005)}]{Stone_2005}
\bibinfo{author}{\bibfnamefont{N.~J.} \bibnamefont{Stone}},
  \bibinfo{journal}{At. Data Nucl. Data Tables} \textbf{\bibinfo{volume}{90}},
  \bibinfo{pages}{75–176} (\bibinfo{year}{2005}).

\bibitem[{\citenamefont{Chasman et~al.}(1977)\citenamefont{Chasman, Ahmad,
  Friedman, and Erskine}}]{Chasman1977}
\bibinfo{author}{\bibfnamefont{R.~R.} \bibnamefont{Chasman}},
  \bibinfo{author}{\bibfnamefont{I.}~\bibnamefont{Ahmad}},
  \bibinfo{author}{\bibfnamefont{A.~M.} \bibnamefont{Friedman}},
  \bibnamefont{and} \bibinfo{author}{\bibfnamefont{J.~R.}
  \bibnamefont{Erskine}}, \bibinfo{journal}{Rev. Mod. Phys.}
  \textbf{\bibinfo{volume}{49}}, \bibinfo{pages}{833} (\bibinfo{year}{1977}),
  \urlprefix\url{https://link.aps.org/doi/10.1103/RevModPhys.49.833}.

\bibitem[{\citenamefont{Engel et~al.}(1975)\citenamefont{Engel, Brink, Goeke,
  Krieger, and Vautherin}}]{Engel75}
\bibinfo{author}{\bibfnamefont{J.}~\bibnamefont{Engel}},
  \bibinfo{author}{\bibfnamefont{D.}~\bibnamefont{Brink}},
  \bibinfo{author}{\bibfnamefont{K.}~\bibnamefont{Goeke}},
  \bibinfo{author}{\bibfnamefont{S.~J.} \bibnamefont{Krieger}},
  \bibnamefont{and}
  \bibinfo{author}{\bibfnamefont{D.}~\bibnamefont{Vautherin}},
  \bibinfo{journal}{Nucl. Phys. A} \textbf{\bibinfo{volume}{249}},
  \bibinfo{pages}{215} (\bibinfo{year}{1975}).

\bibitem[{\citenamefont{Hellemans et~al.}(2012)\citenamefont{Hellemans, Heenen,
  and Bender}}]{Hellemans12}
\bibinfo{author}{\bibfnamefont{V.}~\bibnamefont{Hellemans}},
  \bibinfo{author}{\bibfnamefont{P.-H.} \bibnamefont{Heenen}},
  \bibnamefont{and} \bibinfo{author}{\bibfnamefont{M.}~\bibnamefont{Bender}},
  \bibinfo{journal}{Phys. Rev. C} \textbf{\bibinfo{volume}{85}},
  \bibinfo{pages}{014326} (\bibinfo{year}{2012}).

\bibitem[{\citenamefont{Ring and Schuck}(1980)}]{RS80}
\bibinfo{author}{\bibfnamefont{P.}~\bibnamefont{Ring}} \bibnamefont{and}
  \bibinfo{author}{\bibfnamefont{P.}~\bibnamefont{Schuck}},
  \emph{\bibinfo{title}{The Nuclear Many-Body Problem}}
  (\bibinfo{publisher}{Springer Verlag, New York}, \bibinfo{year}{1980}).

\bibitem[{\citenamefont{Chen et~al.}(2021)\citenamefont{Chen, Li, Dobaczewski,
  and Nazarewicz}}]{Chen21}
\bibinfo{author}{\bibfnamefont{M.}~\bibnamefont{Chen}},
  \bibinfo{author}{\bibfnamefont{T.}~\bibnamefont{Li}},
  \bibinfo{author}{\bibfnamefont{J.}~\bibnamefont{Dobaczewski}},
  \bibnamefont{and}
  \bibinfo{author}{\bibfnamefont{W.}~\bibnamefont{Nazarewicz}},
  \bibinfo{journal}{Phys. Rev. C} \textbf{\bibinfo{volume}{103}},
  \bibinfo{pages}{034303} (\bibinfo{year}{2021}).

\bibitem[{\citenamefont{Nomura et~al.}(2021)\citenamefont{Nomura, Lotina,
  Nik\v{s}i\'{c}, and Vretenar}}]{Nom21}
\bibinfo{author}{\bibfnamefont{K.}~\bibnamefont{Nomura}},
  \bibinfo{author}{\bibfnamefont{L.}~\bibnamefont{Lotina}},
  \bibinfo{author}{\bibfnamefont{T.}~\bibnamefont{Nik\v{s}i\'{c}}},
  \bibnamefont{and} \bibinfo{author}{\bibfnamefont{D.}~\bibnamefont{Vretenar}},
  \bibinfo{journal}{Phys. Rev. C} \textbf{\bibinfo{volume}{103}},
  \bibinfo{pages}{054301} (\bibinfo{year}{2021}).

\bibitem[{\citenamefont{Dutta et~al.}(1991)}]{Dutta91}
\bibinfo{author}{\bibfnamefont{S.~B.} \bibnamefont{Dutta}}
  \bibnamefont{et~al.}, \bibinfo{journal}{Z. Phys. A}
  \textbf{\bibinfo{volume}{341}}, \bibinfo{pages}{39} (\bibinfo{year}{1991}).

\bibitem[{\citenamefont{Leander and Chen}(1988)}]{LC88}
\bibinfo{author}{\bibfnamefont{G.~A.} \bibnamefont{Leander}} \bibnamefont{and}
  \bibinfo{author}{\bibfnamefont{Y.~S.} \bibnamefont{Chen}},
  \bibinfo{journal}{Phys. Rev. C} \textbf{\bibinfo{volume}{37}},
  \bibinfo{pages}{2744} (\bibinfo{year}{1988}).

\end{thebibliography}

\end{document}